%% file: main.tex
\PassOptionsToPackage{table,dvipsnames}{xcolor}
\documentclass{fairmeta}

\title{When History Misleads: Asymmetric Margin Supervision for Instruction-Guided LLM\newline Generative Recommendation}

\author[1,2,*]{Ming~Yin}
\author[2]{Yuhan~Yang}
\author[2]{Chen~Chen}
\author[2]{Xinyu~Lin}
\author[2]{Wentao~Shi}
\author[2]{Fangcong~Yin}
\author[2]{Chaofei~Yang}
\author[2]{Chao~Yang}
\author[2]{Jiyan~Yang}
\author[2]{Hui~Zhang}
\author[2]{Ning~Jiang}
\author[1]{Yiran~Chen}
\author[2]{Qifan~Wang}

\affiliation[1]{Duke University}
\affiliation[2]{Meta}

\contribution[*]{Work done during internship at Meta}

\correspondence{Ming Yin at \email{mingyin@meta.com}}

\input{math_commands.tex}

\usepackage{tikz}
\usepackage{pgfplots}
\pgfplotsset{compat=1.18}
\usepackage{url}
\usepackage{amsfonts}
\usepackage{bbm}

\usepackage{nicefrac}       
\tcbuselibrary{listings,breakable}
\usepackage{fontawesome5}
\tcbuselibrary{skins,breakable,raster}
\usepackage{amsmath, amsthm, amssymb}
\usepackage{colortbl}
\usepackage{collcell}
\usepackage{arydshln}
\usepackage{tabularx}
\usepackage{wrapfig}
\usepackage{pifont}
\usepackage{capt-of}
\usepackage{mathtools}
\usepackage{enumitem}
\setlist[itemize]{leftmargin=*,itemsep=4.5pt,parsep=0pt,topsep=2pt,partopsep=0pt}
\usepackage{pythonhighlight}

\definecolor{linkcolor}{HTML}{ED1C24}

\definecolor{color1}{HTML}{ffdebf}
\definecolor{color2}{HTML}{ffefe0}
\definecolor{color3}{HTML}{E6ECE3}
\definecolor{color4}{HTML}{feeafa}
\definecolor{color5}{HTML}{dee2ff}
\definecolor{color6}{HTML}{ffc0c3}
\definecolor{color7}{HTML}{00ffff}
\definecolor{dash1}{HTML}{db8a87}
\definecolor{dash2}{HTML}{ded6e6}
\newcolumntype{b}{>{\columncolor{gray!10}}c}
\definecolor{back_color}{gray}{0.95}
\definecolor{easy_color}{HTML}{2e4057}
\definecolor{medium_color}{HTML}{083d77}
\definecolor{hard_color}{HTML}{DA4167}
\definecolor{traj_color}{HTML}{3092BF}

\renewcommand{\cite}{\citep}

\definecolor{tab:orange}{HTML}{ff7f0d}
\definecolor{tab:blue}{HTML}{2077b4}

\definecolor{xinyugreen}{RGB}{0,100,0}

\input{sections/0_abstract}

\begin{document}

\maketitle

\input{sections/1_introduction}
\input{sections/promblem}
\input{sections/3_method}
\input{sections/4_experiments}
\input{sections/2_related}
\input{sections/6_conclusion}

\clearpage
\newpage
\bibliographystyle{assets/plainnat}
\bibliography{references}

\clearpage
\newpage
\beginappendix
\input{sections/appendix}

\end{document}

%% file: math_commands.tex
\usepackage{amsmath,amsfonts,bm}

\def\eqref#1{equation~\ref{#1}}

\def\1{\bm{1}}

\DeclareMathAlphabet{\mathsfit}{\encodingdefault}{\sfdefault}{m}{sl}
\SetMathAlphabet{\mathsfit}{bold}{\encodingdefault}{\sfdefault}{bx}{n}



%% file: sections/0_abstract.tex
\abstract{
In instruction-guided generative recommendation, LLM-based recommenders
need to balance two goals: responding to the user's current request and
aligning with the preferences in their interaction history. When the two
conflict, history events can override the request. We show that turning the effect of
individual history events into supervision faces two obstacles. First,
the events that most influence a recommendation are not necessarily the
ones that support the target item. Second, removing a misleading event
can raise the target's score but a competing item's score even more, so a
higher target score alone does not guarantee a better ranking.
We propose Asymmetric Intervention-Guided Margin Supervision (AIMS), which
converts the effect of removing individual history events into ranking
supervision. For training requests already ranked correctly, a frozen
reference model identifies request-specific deletions that improve both
the target's score and its margin over a competitor near the
recommendation cutoff. These margins serve as training targets, while the
complete history is retained as input. Training combines cross-entropy with an asymmetric auxiliary loss that penalizes margin shortfalls and routes its gradient only through the competitor score.
Inference is unchanged, requiring no history editing or deletion search.
Across six LLM backbones on an industrial dataset and two public
benchmarks, AIMS improves Recall and NDCG over strong
baselines. Ablations support request-specific margins and asymmetric
supervision, and the selected deletions preferentially remove
constraint-violating history.
}

%% file: sections/1_introduction.tex
\section{Introduction}
\label{sec:intro}

Generative recommenders retrieve and rank items by decoding item identifiers autoregressively. In
the semantic-identifier (SID) design, each identifier is a short sequence of discrete
codes derived from item content~\citep{rajput2023generative}. Recent systems build on a pre-trained large language model
(LLM)~\citep{geng2022p5,bao2023tallrec,zheng2024lcrec}, whose ability to
interpret natural language enables \emph{instruction-guided
recommendation}~\citep{zhang2025instructrec,lu2024controllable,xu2025iagent},
the setting of this work. Given a natural-language request that states
the user's current intent and an interaction history that records past
behavior, the model should retrieve items that satisfy the request and
match the user's preferences. Ideally, the two inputs play different roles: the
request constrains what is acceptable now, and the history personalizes the choice among acceptable
items.

\looseness=-1
In practice, however, history events that conflict with the current request can override it.
Suppose a user who watched a cooking video of braised pork ribs asks for a quick
vegetarian weeknight dinner, under 30 minutes (Figure~\ref{fig:aims}). The video suggests a taste for slow-cooked
meat, so the model ranks a three-hour beef stew higher than it would without the video, although the stew is neither
vegetarian nor quick. Meanwhile, the same video pushes down the 20-minute chickpea pasta the user wants (the \emph{target}).

Existing approaches improve the use of interaction histories through
counterfactual training and context-bias
mitigation~\citep{zhang2024causality,wang2026context}, edit or select
history events before supplying them to the
model~\citep{zhang2021causerec,lin2023steam,lin2024rella}, or reduce
output bias through preference
optimization~\citep{gao2025sprec}. Editing a history can reduce the
effect of an unhelpful event, but may also remove information useful
for personalizing the recommendation. We instead use the
request-dependent effects of single history events to define a score
gap between the target and a competitor near the recommendation cutoff,
which the model learns from the complete history.

Using single-event interventions as supervision requires knowing which ones help, and
an audit of frozen recommenders (Section~\ref{sec:ium}, Figure~\ref{fig:instruction_ium}) reveals two
obstacles. First, the magnitude of an event's influence on the
prediction, a label-free quantity, does not determine whether the event
supports or suppresses the target: the most influential event in a
history is not reliably more likely than a random one to support the
target. We call
this separation between prediction influence and signed target support
\emph{Influence--Utility Misalignment} (IUM). Second, a higher target score does not necessarily improve the
recommendation: a deletion can raise the target score while raising
that of a competing item near the recommendation cutoff even more,
reducing their score gap (the target--competitor margin). These findings motivate two design
choices. We evaluate interventions offline, where the observed training target,
unavailable at inference, is known and signed support can be measured
directly. We accept an
intervention only if it improves both the target score and its margin
over the boundary competitor, treating this margin as a surrogate for,
not a guarantee of, the decoded decision. Selecting deletions by
label-free influence, or by target uplift instead of margin gain, yields
fewer eligible deletions and lower Recall@10 (Section~\ref{sec:ablations}).

We therefore propose \emph{Asymmetric Intervention-Guided Margin
Supervision} (\textbf{AIMS}), in which a history deletion defines the
supervision, not the training input. A frozen reference model evaluates
single-event deletions, accepts a deletion only if it improves both the
target score and the target--competitor margin, and caches the largest
accepted margin as a request-specific \emph{counterfactual margin}. A student initialized from the same
checkpoint learns this margin from the complete
history. Cross-entropy (CE) supervises the observed target on all
training requests, and an auxiliary loss, \emph{asymmetric competitor
suppression} (ACS), penalizes the student's margin when it falls below
the counterfactual margin by more than a tolerance. This loss is
asymmetric in its gradient: since the target is already supervised
directly, its score enters the forward margin but is detached, so the
auxiliary gradient passes only through the competitor score. It is also selective in scope: ACS applies only to initially correct training requests, to help protect them from displacement during continued training, yet the gains also extend to initially missed requests (Section~\ref{sec:ablations}). At inference, AIMS uses the original decoder on the complete input, with
no reference-model evaluation or deletion search.

In summary, our main contributions are as follows:
\begin{itemize}
    \item \textbf{Diagnosis.} We document Influence--Utility Misalignment: a history event's influence on the prediction does not determine its signed support for the target. We also show that target-score gains alone do not determine the decoded ranking.
  \item \textbf{Method.} AIMS converts instruction-conditioned history deletions into counterfactual margin targets and penalizes margin shortfalls on complete histories through an asymmetric auxiliary loss.
    \item \textbf{Evaluation.} On an industrial dataset and two public benchmarks, AIMS improves Recall and NDCG over strong baselines. Matched ablations support request-specific margins, margin-based deletion selection, complete-history training, and ACS. On constraint-bearing training requests, the selected deletions remove violating history events more often than expected by chance. A history-dropout control does not reproduce the gains.
\end{itemize}

%% file: sections/promblem.tex
\section{Diagnosing Instruction-Conditioned History Use}
\label{sec:ium}

We study instruction-guided generative recommendation, where a
natural-language request $q$ specifies the user's current intent and an
interaction history $\mathcal H=(h_1,\ldots,h_n)$ records prior behavior.
Each item is represented by a semantic identifier (SID)
$\mathbf v=(v_1,\ldots,v_L)$, with sequence score
\begin{equation}
    s_\theta(\mathbf v\mid q,V)
    =\sum_{\ell=1}^{L}
      \log p_\theta(v_\ell\mid q,V,v_{<\ell}),
    \label{eq:score}
\end{equation}
where $V$ denotes a history view. SID-constrained beam search returns a list of SIDs ranked by these scores.
Each evaluated request has one observed target $y$ and is a hit if $y$
appears in the top $R$ of that list. A single-event history intervention either replaces one event with
a placeholder (masking) or removes it (deletion), while leaving the
request and all other history events unchanged.

In the running example (Figure~\ref{fig:aims}), masking the pork-ribs
video changes the model's prediction, but the size of this change does
not reveal whether the video supports the pasta target; in general, a
deletion that raises the target's score can raise the boundary
competitor's score even more.
We therefore separate an event's influence on the prediction, its signed
utility for the target, and the resulting margin
change, and show that utility depends on the request
(Section~\ref{sec:ium_utility}), that influence does not determine it
(Section~\ref{sec:ium_influence}), and that target-score gains do not
guarantee a top-$R$ hit (Section~\ref{sec:ium_decision}).

\begin{figure}[t]
    \centering

    \includegraphics[width=0.9\linewidth]{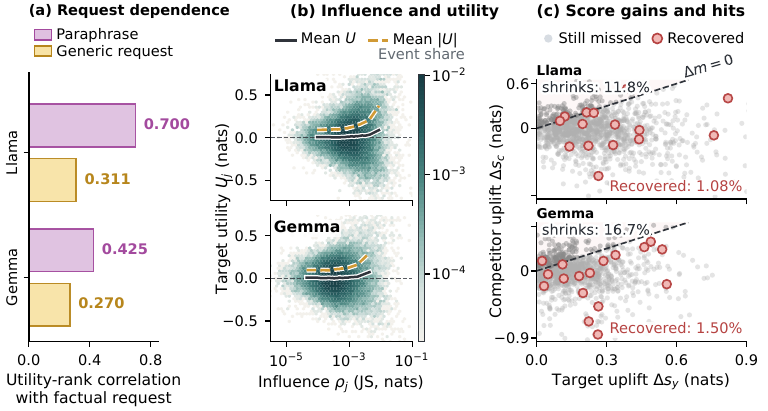}
    \caption{\looseness=-1 \textbf{Diagnosing instruction-conditioned history use} with the frozen SFT checkpoints of Llama-3.1-8B and Gemma-3-12B.
    \textbf{(a,b)}~Single-event masking on 1,600 training requests and 47,972 history events.
    \textbf{(a)}~Within-history Spearman correlation of utility ranks between the factual request and a paraphrase or generic request.
    \textbf{(b)}~Influence $\rho_j$ versus target utility $U_j$ under the factual request; color shows event share (log scale), and solid and dashed curves show mean $U_j$ and mean $|U_j|$ per influence bin.
    \textbf{(c)}~Target versus competitor uplift for the largest-uplift deletion on initially missed requests whose uplift exceeds $10^{-4}$ nats; above the diagonal the margin decreases, and highlighted markers are requests that become top-10 hits after re-decoding.}
    \label{fig:instruction_ium}
    \vspace{-5mm}
\end{figure}

\subsection{Request dependence of utility}
\label{sec:ium_utility}
Let $\theta_0$ denote a checkpoint obtained by supervised fine-tuning
(SFT) a pre-trained LLM to predict target SIDs given requests and
histories. We keep this checkpoint frozen throughout the audits. For event $h_j$, let $V_j$ be the history view obtained
by intervening on that event while leaving all others unchanged. We
define the event's \emph{utility} for candidate $\mathbf v$ as
\begin{equation}
    u_{j,\mathbf v}(q)
    =s_{\theta_0}(\mathbf v\mid q,\mathcal H)
     -s_{\theta_0}(\mathbf v\mid q,V_j).
    \label{eq:utility}
\end{equation}
Positive utility means that retaining the event supports the candidate's
score; negative utility means that the intervention raises it. For the
observed target, we write $U_j(q)=u_{j,\mathbf y}(q)$.

We first test whether utility depends on the request, with $V_j$ masking $h_j$. On the industrial audit set, we compare three conditions per request: the \emph{factual} request, an LLM-generated, meaning-preserving \emph{paraphrase} that controls for wording, and a \emph{generic request} that omits the specific intent. Histories, targets, and masked events are fixed, and full and masked scores are compared within each condition.

For both backbones, paraphrases preserve the factual utility ordering
better than generic requests (Figure~\ref{fig:instruction_ium}(a)).
The utility-sign disagreement rate with the factual condition is
higher for generic requests than for paraphrases, by $17.10$ and $4.65$
points for Llama and Gemma. Thus, \textit{an event's
utility for a fixed target depends on the request, even with the history and reference model fixed}.

\subsection{Influence--utility misalignment}
\label{sec:ium_influence}

Identifying events that support the observed target requires knowing
that target, which is unavailable at inference. We therefore test
whether prediction sensitivity, a label-free signal, can identify them
instead. We define the \emph{influence} $\rho_j(q)$ of $h_j$ as the
Jensen--Shannon divergence between the full- and masked-history
first-token distributions, each normalized over valid first-level SID
codes.

\looseness=-1
Influence is more strongly associated with utility magnitude than with
signed utility (Figure~\ref{fig:instruction_ium}(b)). Under factual
requests, its Spearman correlations with $|U_j|$ are $0.300$ for Llama
and $0.167$ for Gemma, compared with $0.045$ and $0.002$ for $U_j$. Influence could still select supportive events, those whose target utility exceeds an audit threshold of $10^{-4}$ nats, if the most influential events tended to support the target. Yet, compared with uniform selection within each history, maximum-influence selection raises the supportive-event rate by only 0.24 and 0.42 points for Llama and Gemma under factual requests, with both 95\% intervals including zero (Appendix~\ref{app:proxies}), and the point estimates turn negative under generic requests ($-3.94$ and $-2.95$).

We call this separation between prediction sensitivity and signed target support Influence--Utility Misalignment (IUM). A signed attribution, the utility $u_{j,\hat{\mathbf{y}}}$ for the top-1 prediction $\hat{\mathbf{y}}$ of $\theta_0$, does not resolve it. When $\hat{\mathbf{y}}$ is not the target, $u_{j,\hat{\mathbf{y}}}$ is weakly anti-correlated with $U_j$ (Appendix~\ref{app:proxies}). Offline, however, the observed target makes this signed score change measurable, so it can serve as supervision (Section~\ref{sec:method}).
\vspace{-1mm}
\subsection{From target scores to margins and decisions}
\label{sec:ium_decision}
\vspace{-1mm}
Even with the observed target, a higher target score does not
necessarily improve the recommendation, because the same intervention
can also raise competing items' scores. For each request, we fix a
competitor $c$ at the top-$R$ boundary of the initial full-history beam:
the first excluded candidate at rank $R+1$ if the target is already in
the top-$R$ list, and the candidate at rank $R$ otherwise. Keeping its
identity fixed across history views, we define
\begin{align}
    m_\theta(\mathbf y,c\mid q,V)
    &=s_\theta(\mathbf y\mid q,V)-s_\theta(c\mid q,V),
    \label{eq:margin}\\
    \Delta_j m_{\theta_0}
    &=\Delta_j s_{\mathbf y}-\Delta_j s_c
      =-u_{j,\mathbf y}(q)+u_{j,c}(q),
    \label{eq:mask_margin}
\end{align}
where $\Delta_j$ denotes the intervened-view value minus the
full-history value at $\theta_0$. The margin improves only when the
target uplift exceeds the competitor uplift. The same decomposition
applies to score changes induced by continued training.

The decoder audit deletes each history event of an initially missed
request in turn ($R=10$), with the deletion operator of AIMS, and keeps
the deletion with the largest target uplift if this uplift exceeds the
audit threshold (Figure~\ref{fig:instruction_ium}(c)). We then measure its
effect on the fixed competitor and the margin and re-decode the
view. Despite the target gain, the margin decreases for
$11.8\%$ (Llama) and $16.7\%$ (Gemma) of the requests, and re-decoding
recovers only $1.08\%$ and $1.50\%$ of them as top-10 hits.

A fixed-competitor margin is pairwise: top-$R$
membership also depends on other candidates and the beam-search output,
so the target can enter the
decoded list even when this margin decreases (recoveries above the
diagonal in Figure~\ref{fig:instruction_ium}(c)). We therefore
check both target uplift and margin gain when constructing
supervision, with the margin as a surrogate for the decoded decision.

%% file: sections/3_method.tex
\section{Asymmetric Intervention-Guided Margin Supervision}
\label{sec:method}

\begin{figure}[t]
    \centering
    \includegraphics[width=\linewidth]{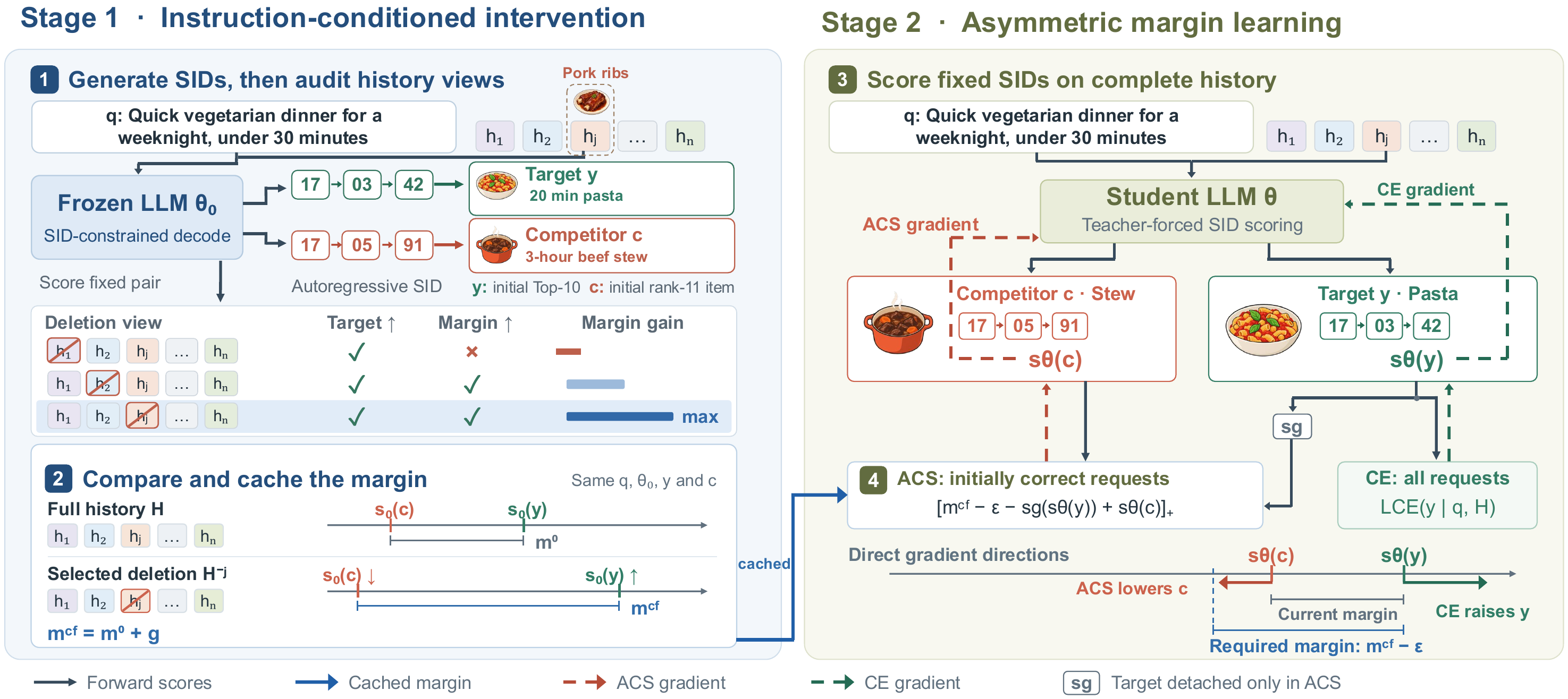}
    \caption{\textbf{Overview of AIMS with deletion budget $K=1$.}
    \textbf{Left:} for an initially correct request, the frozen reference
    model scores the target $\mathbf y$ and the competitor $c$ under the
    complete history and each deletion view, accepts deletions that
    raise both the target score and the margin, and caches the largest
    accepted margin as $m^{\mathrm{cf}}$ (the complete-history margin if
    none qualifies). \textbf{Right:} the student sees the complete input; CE supervises the target on all requests, and on
    initially correct requests ACS penalizes a shortfall below
    $m^{\mathrm{cf}}-\varepsilon$ through the competitor score, with the
    target detached ($\operatorname{sg}$).}
    \label{fig:aims}
    \vspace{-3mm}
\end{figure}

AIMS turns instruction-conditioned history deletions into asymmetric
margin supervision (Figure~\ref{fig:aims}). Section~\ref{sec:clean_teacher}
builds a counterfactual margin for each initially correct training
request, and Section~\ref{sec:acs} trains the student to reach it from
the complete history.

\subsection{Counterfactual margins from history deletions}
\label{sec:clean_teacher}
The frozen SFT checkpoint $\theta_0$ serves as the reference model and initializes the student. Continued CE training recovers
some initially missed targets but also loses some initial top-10
hits (Section~\ref{sec:ablations}). AIMS therefore adds supervision on the \emph{protection set}
$\mathcal P$: the single-target training requests whose target appears
in the initial complete-history top-$R$ list ($R=10$). For
$i\in\mathcal P$, let $q_i$, $\mathcal H_i$, and $\mathbf y_i$ denote its
request, history, and target SID, and let $c_i$ be the competitor, the candidate at rank $R+1$ of the
initial decoded list. Its identity stays fixed across history views and
throughout training. We write
  \begin{equation}
      m_\theta(i;V)=s_\theta(\mathbf y_i\mid q_i,V)-s_\theta(c_i\mid q_i,V),
      \qquad m_i^0=m_{\theta_0}(i;\mathcal H_i),
      \label{eq:aims_margin}
  \end{equation}
where both scores are the full-sequence SID log-probabilities of
Eq.~(\ref{eq:score}), computed by teacher forcing, and the reference and
the student use the same scoring procedure.

We formulate the intervention with a deletion budget $K$. For a history
of length $n_i$, let $\mathcal C_i^{(K)}$ be the collection of evaluated
index subsets $S\subseteq\{1,\ldots,n_i\}$ with $1\leq|S|\leq K$; for the
default $K=1$ it contains every single-event deletion. Deleting the
events in $S$ gives the view $\mathcal H_i^{-S}$, which keeps the
request and the order of the remaining events. For each view we measure
the target uplift and the margin gain relative to the complete history,
\begin{equation}
    \Delta_{y,i}(S)
    =s_{\theta_0}(\mathbf y_i\mid q_i,\mathcal H_i^{-S})
     -s_{\theta_0}(\mathbf y_i\mid q_i,\mathcal H_i),
    \qquad
    \Delta_{m,i}(S)
    =m_{\theta_0}(i;\mathcal H_i^{-S})-m_i^0,
    \label{eq:clean_gains}
\end{equation}
and accept the views that improve both,
\begin{equation}
    \mathcal E_i^{(K)}
    =\bigl\{\,S\in\mathcal C_i^{(K)}:\
      \Delta_{y,i}(S)>\tau,\ \Delta_{m,i}(S)>\tau\,\bigr\},
    \label{eq:aims_eligible}
\end{equation}
where $\tau>0$ excludes negligible changes (Appendix~\ref{app:hparams}).
The target-uplift condition rejects a wider margin while the target's
own score falls, and the margin-gain condition rejects a target gain
that the competitor matches or exceeds
(Figure~\ref{fig:aims}, left). An accepted deletion improves the
reference comparison under $q_i$; the removed events may still be
informative under other requests.

Among accepted views we take the largest margin gain and define the
counterfactual margin
\begin{equation}
    g_i=\begin{cases}
        \displaystyle\max_{S\in\mathcal E_i^{(K)}}\Delta_{m,i}(S),
            & \mathcal E_i^{(K)}\neq\varnothing,\\[2pt]
        0, & \mathcal E_i^{(K)}=\varnothing,
    \end{cases}
    \qquad
    m_i^{\mathrm{cf}}=m_i^0+g_i.
    \label{eq:clean_teacher}
\end{equation}
Without an eligible deletion, $m_i^{\mathrm{cf}}=m_i^0$ and the request
stays in $\mathcal P$, so eligible deletions strengthen the margin target
without changing the protection set. Selection uses training labels and
runs once, offline; the protection set, competitors, and counterfactual
margins are then fixed for training. Deleting up to three events or
averaging ACS over up to five competitors changes Recall@10 only
marginally, the former with up to $2.9\times$ as many scoring views,
whereas taking the maximum over competitors degrades as the set grows
(Appendix~\ref{app:sensitivity}); we therefore use $K=1$ and one
competitor.

\subsection{Learning the counterfactual margin on the complete history}
\label{sec:acs}

The student learns the cached margin $m_i^{\mathrm{cf}}$ from the
complete input $(q_i,\mathcal H_i)$, the history available at inference:
deletions define the margin target, not the input.

Write $s_\theta(\mathbf v\mid i)=s_\theta(\mathbf v\mid q_i,\mathcal H_i)$
for the student's complete-history score, and let
$\operatorname{sg}(\cdot)$ return its argument in the forward pass and
block its gradient. For a minibatch $\mathcal B$ we minimize
\begin{equation}
\begin{aligned}
    \ell_{\mathrm{ACS}}(i;\theta)
    &=\Bigl[m_i^{\mathrm{cf}}-\varepsilon
       -\operatorname{sg}\!\bigl(s_\theta(\mathbf y_i\mid i)\bigr)
       +s_\theta(c_i\mid i)\Bigr]_+,\\
       \mathcal L(\theta;\mathcal B)
    &=\mathcal L_{\mathrm{CE}}(\theta;\mathcal B)
      +\frac{\lambda}{|\mathcal B|}
       \sum_{i\in\mathcal B\cap\mathcal P}
       \ell_{\mathrm{ACS}}(i;\theta),
    \qquad
    \mathcal L_{\mathrm{CE}}(\theta;\mathcal B)
    =-\frac{1}{L|\mathcal B|}\sum_{i\in\mathcal B}s_\theta(\mathbf y_i\mid i),
\end{aligned}
\label{eq:acs}
\end{equation}
where $\mathcal L_{\mathrm{CE}}$ is the token-level cross-entropy over the
$L=3$ SID tokens, shared with Continued CE, $[z]_+=\max\{z,0\}$, $\lambda$ weights the auxiliary
loss, and $\varepsilon\geq0$ tolerates a small shortfall.

Because CE already supervises the target, ACS routes its gradient only
through the competitor: the detached target score still affects hinge
activation, and when
$m_\theta(i;\mathcal H_i)<m_i^{\mathrm{cf}}-\varepsilon$
(Figure~\ref{fig:aims}, right, step~4) the gradient is
$\nabla_\theta s_\theta(c_i\mid i)$, and zero otherwise.
By default, ACS differentiates the full competitor score, including any
SID prefix shared with the target, and shared parameters can still
change both scores; Section~\ref{sec:ablations}
compares target-only, symmetric, and prefix-safe routing.

At initialization the student equals the reference, so
$\ell_{\mathrm{ACS}}(i;\theta_0)=[g_i-\varepsilon]_+$: ACS starts active
for requests with $g_i>\varepsilon$, which covers most protected requests
(Appendix~\ref{app:protection}), and for $g_i=0$ it activates only if the
margin falls more than $\varepsilon$ below $m_i^0$. Inference is
unchanged.

%% file: sections/4_experiments.tex
\begin{table*}[!b]
  \centering
  \caption{\textbf{Performance (\%) on the industrial dataset.} Bold: best
    per model. AIMS significantly exceeds the strongest baseline in
    Recall@10 for every model (paired user-cluster bootstrap, $p<0.05$).}
  \label{tab:internal-main-results}

  \fontsize{8}{9.2}\selectfont
\setlength{\tabcolsep}{8.88pt}
\renewcommand{\arraystretch}{0.92}
  \setlength{\aboverulesep}{0.20ex}
  \setlength{\belowrulesep}{0.20ex}
  \setlength{\cmidrulesep}{0.10ex}

  \begin{tabular}{
    ll *{7}{r}
  }
    \toprule
    Model & Method
      & \multicolumn{4}{c}{Recall (\%) $\uparrow$}
      & \multicolumn{3}{c}{NDCG (\%) $\uparrow$} \\
    \cmidrule(lr){3-6}\cmidrule(lr){7-9}
      & & R@1 & R@5 & R@10 & R@20
        & N@5 & N@10 & N@20 \\
    \midrule

    \multirow{6}{*}{Llama-3.1-8B}
      & Continued CE
      & 2.475 & 5.315 & 6.902 & 8.773
      & 3.949 & 4.460 & 4.934 \\

      & CFT
      & 2.501 & 5.322 & 6.939 & 8.776
      & 3.961 & 4.451 & 4.942 \\

      & LETTER (RG)
      & 2.503 & 5.392 & 7.136 & 9.029
      & 4.018 & 4.549 & 5.057 \\

      & LTRGR
      & 2.511 & 5.389 & 7.092 & 9.041
      & 3.998 & 4.514 & 5.016 \\

      & S-DPO
      & 2.489 & 5.306 & 6.977 & 8.853
      & 3.953 & 4.512 & 4.965 \\

      \rowcolor{gray!8}\cellcolor{white}
      & \textbf{AIMS (ours)}
      & \textbf{2.547} & \textbf{5.497}
      & \textbf{7.223} & \textbf{9.097}
      & \textbf{4.074} & \textbf{4.629} & \textbf{5.102} \\
    \midrule

    \multirow{6}{*}{Llama-3.1-70B}
      & Continued CE
      & 4.537 & 8.070 & 9.990 & 12.043
      & 6.382 & 7.006 & 7.524 \\

      & CFT
      & 4.475 & 8.002 & 9.974 & 11.952
      & 6.328 & 6.968 & 7.469 \\

      & LETTER (RG)
      & \textbf{5.056} & 9.000 & 10.870 & 13.017
      & 7.103 & 7.669 & 8.191 \\

      & LTRGR
      & 5.013 & 8.839 & 10.340 & 12.829
      & 6.977 & 7.432 & 8.025 \\

      & S-DPO
      & 4.962 & 8.745 & 10.274 & 12.643
      & 6.912 & 7.375 & 7.951 \\

      \rowcolor{gray!8}\cellcolor{white}
      & \textbf{AIMS (ours)}
      & 5.053 & \textbf{9.031}
      & \textbf{10.983} & \textbf{13.049}
      & \textbf{7.144} & \textbf{7.772} & \textbf{8.295} \\
    \midrule

    \multirow{6}{*}{Gemma-3-4B}
      & Continued CE
      & 1.180 & 2.683 & 3.681 & 4.809
      & 1.963 & 2.278 & 2.571 \\

      & CFT
      & 1.225 & 2.801 & 3.808 & 4.993
      & 2.041 & 2.366 & 2.667 \\

      & LETTER (RG)
      & 1.282 & 2.865 & 3.824 & 5.040
      & 2.079 & 2.388 & 2.701 \\

      & LTRGR
      & 1.268 & 2.829 & 3.875 & 5.016
      & 2.078 & 2.417 & 2.723 \\

      & S-DPO
      & 1.196 & 2.773 & 3.810 & 4.989
      & 2.073 & 2.387 & 2.682 \\

      \rowcolor{gray!8}\cellcolor{white}
      & \textbf{AIMS (ours)}
      & \textbf{1.295} & \textbf{2.897}
      & \textbf{3.911} & \textbf{5.128}
      & \textbf{2.118} & \textbf{2.446} & \textbf{2.754} \\
    \midrule

    \multirow{6}{*}{Gemma-3-12B}
      & Continued CE
      & 2.862 & 6.050 & 7.945 & 10.040
      & 4.526 & 5.141 & 5.669 \\

      & CFT
      & 2.850 & 6.113 & 7.907 & 10.051
      & 4.550 & 5.130 & 5.671 \\

      & LETTER (RG)
      & 3.142 & 6.541 & 8.496 & 10.725
      & 4.908 & 5.508 & 6.091 \\

      & LTRGR
      & 2.955 & 6.274 & 8.189 & 10.310
      & 4.692 & 5.312 & 5.845 \\

      & S-DPO
      & 3.074 & 6.492 & 8.155 & 10.285
      & 4.707 & 5.277 & 5.849 \\

      \rowcolor{gray!8}\cellcolor{white}
      & \textbf{AIMS (ours)}
      & \textbf{3.181} & \textbf{6.641}
      & \textbf{8.609} & \textbf{10.742}
      & \textbf{4.962} & \textbf{5.598} & \textbf{6.140} \\
    \midrule

    \multirow{6}{*}{Qwen3-8B}
      & Continued CE
      & 2.286 & 4.944 & 6.438 & 8.224
      & 3.658 & 4.143 & 4.595 \\

      & CFT
      & 2.312 & 4.921 & 6.471 & 8.193
      & 3.651 & 4.167 & 4.603 \\

      & LETTER (RG)
      & 2.349 & \textbf{5.098} & 6.635 & 8.468
      & 3.774 & 4.272 & 4.737 \\

      & LTRGR
      & 2.361 & 5.071 & 6.611 & 8.431
      & 3.751 & 4.266 & 4.724 \\

      & S-DPO
      & 2.303 & 5.006 & 6.496 & 8.336
      & 3.711 & 4.196 & 4.654 \\

      \rowcolor{gray!8}\cellcolor{white}
      & \textbf{AIMS (ours)}
      & \textbf{2.389} & 5.097
      & \textbf{6.695} & \textbf{8.534}
      & \textbf{3.803} & \textbf{4.327} & \textbf{4.793} \\
    \midrule

    \multirow{6}{*}{Qwen3-14B}
      & Continued CE
      & 2.665 & 5.681 & 7.354 & 9.321
      & 4.219 & 4.758 & 5.257 \\

      & CFT
      & 2.648 & 5.724 & 7.381 & 9.358
      & 4.236 & 4.771 & 5.281 \\

      & LETTER (RG)
      & 2.842 & 5.986 & 7.706 & 9.728
      & 4.448 & 5.021 & 5.556 \\

      & LTRGR
      & 2.794 & 5.933 & 7.655 & 9.671
      & 4.402 & 4.988 & 5.517 \\

      & S-DPO
      & 2.823 & 5.867 & 7.544 & 9.589
      & 4.388 & 4.908 & 5.442 \\

      \rowcolor{gray!8}\cellcolor{white}
      & \textbf{AIMS (ours)}
      & \textbf{2.901} & \textbf{6.061}
      & \textbf{7.842} & \textbf{9.854}
      & \textbf{4.509} & \textbf{5.119} & \textbf{5.659} \\

    \bottomrule
    \vspace{-6mm}
  \end{tabular}
\end{table*}
\vspace{-3mm}
\section{Experiments}
  \label{sec:experiments}
\vspace{-2mm}
  \subsection{Experimental Setup}
  \label{sec:setup}
\paragraph{Datasets and protocols.}
We use an industrial dataset and two public benchmarks,
Qilin~\citep{chen2025qilin} and KuaiSearch-Lite, the benchmark subset of
the KuaiSearch e-commerce search dataset~\citep{li2026kuaisearch}. The
industrial dataset is sampled from anonymized production logs of a
large-scale content platform; each request is a natural-language user
query, and the target is the item the user clicked. On it, every method
runs SID-constrained beam search over the full catalog. On the public
benchmarks, every method ranks each request's released candidate pool by
the teacher-forced SID log-probability (Eq.~\ref{eq:score}); pools with
fewer than 40 candidates, too small for an informative top-10 cutoff, are
extended with training-split impressions sampled without click labels,
identically for all methods (Appendix~\ref{app:setup}).

\paragraph{Models and baselines.}
On all three datasets we use six backbones from three families:
Llama-3.1-8B/70B,
Gemma-3-4B/12B, and
Qwen3-8B/14B. For each dataset and backbone, all
methods start from the same SFT checkpoint $\theta_0$. \emph{Continued CE}
continues CE-only training from $\theta_0$ and is the matched baseline for
AIMS. We also compare with CFT~\citep{zhang2024causality}, LETTER
(RG)~\citep{wang2024letter}, LTRGR~\citep{li2024ltrgr}, and
S-DPO~\citep{chen2024sdpo}, described in Appendix~\ref{app:setup}. We
report Recall@$k$ and NDCG@$k$ in percent, with differences in percentage
points and significance from paired user-cluster bootstrap tests.
\vspace{-1mm}

\subsection{Main Results}
  \label{sec:main-results}
  \vspace{-2mm}
AIMS achieves the best Recall@10 and NDCG@10 for all six backbones on all
three datasets (Tables~\ref{tab:internal-main-results} and~\ref{tab:public}). Over the
matched Continued CE baseline, it improves Recall@10 in every setting, by
4.0--10.9\% in relative terms. In repeated runs with three training seeds,
it also exceeds both Continued CE and the strongest baseline in every run
(Table~\ref{tab:seeds}).

\begin{table}[!h]
    \vspace{-1mm}
    \centering
    \caption{\textbf{Results on the public benchmarks.} Recall@10 / NDCG@10
    (\%) under candidate-pool ranking (Section~\ref{sec:setup}). Bold marks the best result per column.
    Appendix~\ref{app:cutoffs} reports all cutoffs.}
    \label{tab:public}
    \fontsize{8}{9.4}\selectfont
    \setlength{\tabcolsep}{3.77pt}
    \renewcommand{\arraystretch}{0.94}
    \setlength{\aboverulesep}{0.20ex}
    \setlength{\belowrulesep}{0.20ex}
    \begin{tabular}{l*{12}{r}}
        \toprule
        & \multicolumn{2}{c}{Llama-3.1-8B} & \multicolumn{2}{c}{Llama-3.1-70B}
        & \multicolumn{2}{c}{Gemma-3-4B} & \multicolumn{2}{c}{Gemma-3-12B}
        & \multicolumn{2}{c}{Qwen3-8B} & \multicolumn{2}{c}{Qwen3-14B} \\
        \cmidrule(lr){2-3}\cmidrule(lr){4-5}\cmidrule(lr){6-7}
        \cmidrule(lr){8-9}\cmidrule(lr){10-11}\cmidrule(lr){12-13}
        Method & R@10 & N@10 & R@10 & N@10 & R@10 & N@10 & R@10 & N@10 & R@10 & N@10 & R@10 & N@10 \\
        \midrule
        \multicolumn{13}{l}{\emph{Qilin}} \\
        Continued CE  & 35.48 & 22.92 & 43.47 & 28.47 & 30.74 & 19.38 & 36.29 & 23.68 & 37.45 & 24.74 & 39.81 & 26.24 \\
        CFT           & 35.52 & 23.04 & 43.35 & 28.83 & 30.63 & 19.55 & 36.80 & 23.67 & 37.62 & 24.39 & 40.91 & 27.07 \\
        LETTER (RG)   & 36.70 & 23.97 & 45.08 & 30.34 & 31.45 & 19.91 & 37.74 & 24.62 & 38.63 & 25.32 & 41.35 & 27.75 \\
        LTRGR         & 36.13 & 23.45 & 44.30 & 29.11 & 31.16 & 19.97 & 37.09 & 24.15 & 38.41 & 25.12 & 41.42 & 27.43 \\
        S-DPO         & 35.87 & 23.49 & 44.19 & 29.50 & 31.94 & 20.71 & 37.64 & 24.54 & 37.80 & 24.74 & 40.97 & 27.23 \\
        \rowcolor{gray!8}
        \textbf{AIMS (ours)}
            & \textbf{37.61} & \textbf{24.41} & \textbf{45.62} & \textbf{30.90}
            & \textbf{32.79} & \textbf{20.95} & \textbf{38.57} & \textbf{25.10}
            & \textbf{39.10} & \textbf{25.88} & \textbf{42.06} & \textbf{27.92} \\
        \addlinespace[1pt]
        \midrule
        \multicolumn{13}{l}{\emph{KuaiSearch-Lite}} \\
        Continued CE  & 26.36 & 17.38 & 31.80 & 21.31 & 23.11 & 14.91 & 28.04 & 18.74 & 23.94 & 15.70 & 27.15 & 17.88 \\
        CFT           & 26.41 & 17.18 & 32.11 & 21.50 & 23.35 & 15.14 & 27.91 & 18.01 & 24.42 & 15.90 & 27.03 & 17.84 \\
        LETTER (RG)   & 26.87 & 17.58 & 33.43 & 22.26 & 23.86 & 15.59 & 28.80 & 18.71 & 25.40 & 16.71 & 28.13 & 18.56 \\
        LTRGR         & 27.21 & 17.75 & 33.16 & 22.14 & 23.79 & 15.47 & 28.39 & 18.98 & 25.33 & 16.69 & 27.82 & 18.19 \\
        S-DPO         & 26.54 & 17.00 & 32.71 & 21.60 & 23.40 & 14.92 & 28.53 & 18.72 & 25.18 & 16.51 & 28.05 & 18.49 \\
        \rowcolor{gray!8}
        \textbf{AIMS (ours)}
            & \textbf{27.70} & \textbf{18.12} & \textbf{34.02} & \textbf{22.55}
            & \textbf{24.75} & \textbf{16.15} & \textbf{29.97} & \textbf{19.82}
            & \textbf{26.56} & \textbf{17.53} & \textbf{28.78} & \textbf{19.03} \\
        \bottomrule
    \end{tabular}
\end{table}
\vspace{-2mm}
\subsection{Ablation Studies and Analysis}
\label{sec:ablations}
\vspace{-2mm}
Each ablation changes one design choice of AIMS and otherwise keeps its
setup, protection set, and competitors; all use Llama-3.1-8B and
Gemma-3-12B on the industrial dataset and are evaluated on complete
histories.
We report paired values as Llama/Gemma and state
significance where tested (paired user-cluster bootstrap, $p<0.05$);
other differences are point estimates.

\begin{figure}[t]
    \centering
    \includegraphics[width=\linewidth]{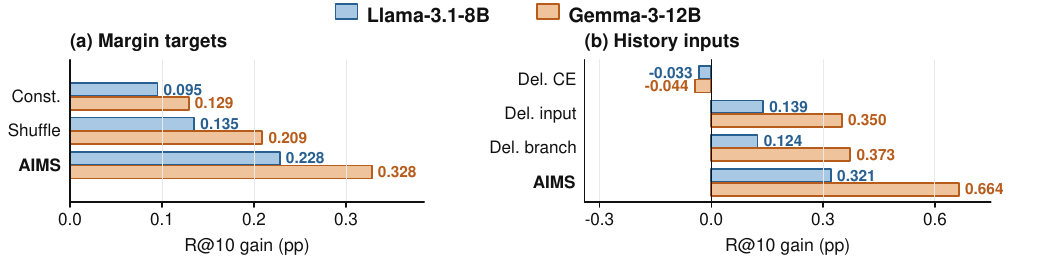}
    \caption{\textbf{Ablation results}, in percentage points (higher is better).
    \textbf{(a)}~Recall@10 gain over R-comp, which uses no intervention gain.
    \textbf{(b)}~Recall@10 gain over Continued CE with the original cached targets; Table~\ref{tab:design-ablations}, group~B, matches initial hinge gaps.
    Const./Shuffle: ConstantGate/Magnitude Shuffle; Del.: Deletion/Deleted.}
    \label{fig:abl}
\end{figure}

\begin{figure}[t]
    \centering
    \captionsetup{skip=5pt}
    \begin{minipage}[t]{0.60\linewidth}
        \vspace{0pt}
        \centering
        \begin{minipage}[t][2.36in][c]{\linewidth}
            \centering
            \fontsize{8.5}{9.55}\selectfont
            \setlength{\tabcolsep}{2.5pt}
            \renewcommand{\arraystretch}{0.92}
            \setlength{\aboverulesep}{0.20ex}
            \setlength{\belowrulesep}{0.20ex}
            \setlength{\cmidrulesep}{0.10ex}
            \begin{tabular*}{\linewidth}{@{\extracolsep{\fill}}lrrrr@{}}
                \toprule
                & \multicolumn{2}{c}{Llama-3.1-8B}
                & \multicolumn{2}{c}{Gemma-3-12B} \\
                \cmidrule(lr){2-3}\cmidrule(lr){4-5}
                Variant & R@10 & N@10 & R@10 & N@10 \\
                \midrule
                Continued CE   & 6.902 & 4.460 & 7.945 & 5.141 \\
                \textbf{AIMS}  & \textbf{7.223} & 4.629 & \textbf{8.609} & 5.598 \\
                \midrule
                \multicolumn{5}{@{}l}{\textit{A. Eligibility, selection, and gain construction}} \\
                Margin-only    & 7.142 & 4.613 & 8.438 & 5.483 \\
                D-free         & 7.051 & 4.545 & 8.224 & 5.317 \\
                D-cal          & 7.097 & 4.572 & 8.321 & 5.391 \\
                Generic-matched & 7.138 & 4.567 & 8.384 & 5.447 \\
                Uplift-first    & 7.091 & 4.561 & 8.426 & 5.478 \\
Influence-first & 7.063 & 4.552 & 8.312 & 5.381 \\
                \addlinespace[1.5pt]
                \multicolumn{5}{@{}l}{\emph{B. History inputs: matched gaps}} \\
                Deleted-branch & 7.106 & 4.571 & 8.340 & 5.412 \\
                Deleted-input  & 7.045 & 4.529 & 8.282 & 5.374 \\
                \addlinespace[1.5pt]
                \multicolumn{5}{@{}l}{\emph{C. Auxiliary gradient routing}} \\
                Target-only    & 6.895 & 4.457 & 7.953 & 5.146 \\
                Symmetric      & 6.955 & 4.488 & 8.253 & 5.362 \\
                Prefix-safe    & 7.212 & \textbf{4.653} & 8.602 & \textbf{5.613} \\
                \bottomrule
            \end{tabular*}
        \end{minipage}
        \captionof{table}{\raggedright\textbf{AIMS design ablations.}
        Recall@10 and NDCG@10 (\%). Group~B uses the gap-matched targets
        $T_i(V)$. Bold marks column maxima.}
        \label{tab:design-ablations}
        \label{tab:matched-gaps}
        \label{tab:gradient-routing}
        \label{tab:eligibility}
    \end{minipage}\hfill
    \begin{minipage}[t]{0.37\linewidth}
        \vspace{0pt}
        \centering
        \begin{minipage}[t][2.36in][c]{\linewidth}
            \centering
            \includegraphics[height=2.32in,width=\linewidth,keepaspectratio]
                {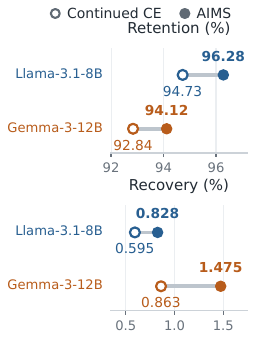}
        \end{minipage}
        \captionof{figure}{\raggedright\textbf{Retention and recovery} (\%),
        conditioned on initial SFT top-10 hits and misses, respectively.}
        \label{fig:retention-recovery}
    \end{minipage}
    \vspace{-3mm}
\end{figure}

\paragraph{Each request's own gain matters.}
Both the intervention gains and their assignment to requests contribute to AIMS. \emph{R-comp}, which removes the gains ($m_i^{\mathrm{cf}}=m_i^0$) and thus only keeps protected margins from falling more than $\varepsilon$ below those of $\theta_0$, still improves Recall@10 over Continued CE by +0.093/+0.336 ($p<0.05$ on both backbones), so the gains account for 71\%/49\% of the improvement of AIMS. \emph{ConstantGate} replaces each positive gain with its mean within the group of initially active or inactive requests, and \emph{Magnitude Shuffle} permutes positive gains within these groups; both keep which requests start active, yet fall between R-comp and AIMS (Figure~\ref{fig:abl}(a)), and AIMS exceeds Shuffle ($p<0.05$ on both backbones), so the assignment itself matters. AIMS also exceeds variants that derive gains from initial-margin difficulty (\emph{D-free}), reassign its gains by that difficulty (\emph{D-cal}), or drop the target-uplift condition (\emph{Margin-only}), and reassigning them by gains computed under a generic request (\emph{Generic-matched}) lowers both metrics, so the specific request matters (Table~\ref{tab:design-ablations}, group~A; Appendix~\ref{app:difficulty-controls}). Inserting a random item at each history position instead yields positive gains about ten times smaller and only weakly correlated with the deletion gains (Appendix~\ref{app:placebo}).

\paragraph{Margin-based deletion selection matters.}
Choosing deletions by the criteria that Section~\ref{sec:ium} cautions against weakens AIMS. \emph{Uplift-first} takes each request's largest-uplift deletion, as audited in Figure~\ref{fig:instruction_ium}(c), and \emph{Influence-first} its most influential one, chosen without labels (Appendix~\ref{app:selection-rules}); each keeps that deletion only if it passes Eq.~(\ref{eq:aims_eligible}). They leave an eligible deletion for only 75.9/66.1\% and 28.8/25.9\% of protected requests, versus 88.2/85.7\% for AIMS, with smaller positive gains (Appendix~\ref{app:placebo}). Both lower Recall@10 from 7.223/8.609 to 7.091/8.426 and 7.063/8.312, and NDCG@10 as well (Table~\ref{tab:design-ablations}, group~A); the drop reflects both the chosen deletion and its targets.

\paragraph{Deletions work better as targets than as training inputs.}
We feed the selected deletion views to the student with CE alone
(\emph{Deletion CE}), in ACS only (\emph{Deleted-branch}), or in both
losses (\emph{Deleted-input}); all use the AIMS deletions, the latter two
also its targets, and requests without an eligible deletion keep
the full history. Deletion CE does not beat Continued CE, and the other
two do but trail AIMS (Figure~\ref{fig:abl}(b)).
As deletion-view ACS starts with a zero hinge under the cached
targets, we also equalize the starting hinge with view-specific targets
$T_i(V)=m_{\theta_0}(i;V)+g_i$; AIMS still leads on both metrics
(Table~\ref{tab:design-ablations}, group~B), though this control
does not equalize later training dynamics.

\paragraph{Competitor-only routing outperforms target-only and symmetric routing.}
The routing variants share the cached targets and, at identical
parameters, the forward hinge and its activation; they differ only in
where the auxiliary gradient flows. Routing it only through the target
(\emph{Target-only}) is on par with Continued CE, and through both scores
(\emph{Symmetric}) falls below AIMS in Recall@10 on both backbones
($p<0.05$, Table~\ref{tab:design-ablations}, group~C), supporting the
separation of CE target learning from competitor suppression. Also
stopping shared-prefix gradients (\emph{Prefix-safe}) gives slightly
lower Recall and slightly higher NDCG.

\begin{figure}[t]
    \centering
    \captionsetup{skip=5pt} 
    \includegraphics[width=0.9\linewidth]{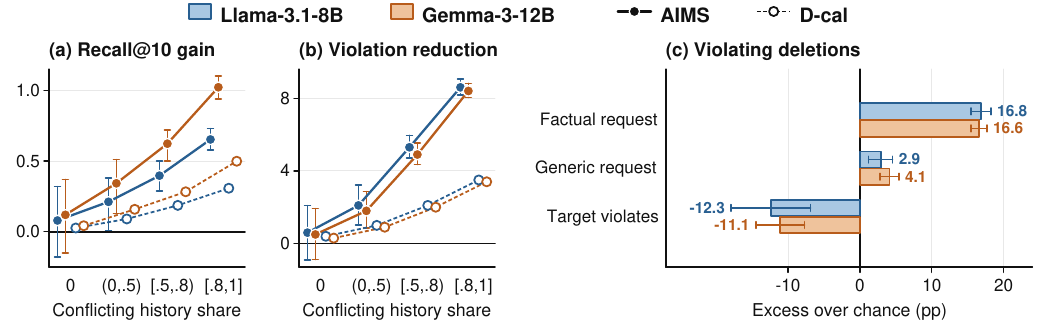}
    \caption{\textbf{Conflict analysis}, in percentage points.
    \textbf{(a,b)}~Recall@10 gain and observed-violation reduction over Continued CE, by the share of evaluable history items that violate the request's constraint (Table~\ref{tab:conflict-dose-response}).
    \textbf{(c)}~Share of AIMS-selected deletions that remove a violating event, minus chance (Table~\ref{tab:sel-labels}).
    Whiskers: 95\% intervals (AIMS only in (a,b)).}
    \vspace{-5mm}
    \label{fig:conflict-analysis}
\end{figure}

\paragraph{Both retention and recovery improve.}
With test requests split by the initial top-10 decision, AIMS raises
both retention (initial hits that stay hits) and
recovery (initial misses that become hits) over Continued CE on both
backbones (all $p<0.05$, Figure~\ref{fig:retention-recovery}), so the
gain does not trade recovery for retention. Higher recovery accounts for
68/85\% of it (Appendix~\ref{app:retention}), even though ACS is applied
only to initially correct training requests. This is consistent with the
lower violation rates on conflict requests (Figure~\ref{fig:conflict-analysis}(b)), suggesting that suppressing boundary competitors may also demote history-aligned items on other requests.

\paragraph{Gains are larger under instruction--history conflict.}
On conflict requests, where at least half of the evaluable history items
violate the request's explicit category constraint, AIMS improves
Recall@10 over Continued CE by +0.558/+0.874 points, versus +0.321/+0.664 on the full test
set, and lowers the observed-violation rate, the share of the ten output slots labeled as confirmed violations,
from 29.0/25.4\% to 21.6/18.3\%. Both effects grow with conflict severity, and so does the advantage over D-cal (Figure~\ref{fig:conflict-analysis}(a,b)). These
comparisons are observational (Appendix~\ref{app:conflict}).
\vspace{-3mm}
\paragraph{Selected deletions remove constraint-violating history.}
On protected training requests with an explicit category constraint, a satisfying target, and a mixed history (Appendix~\ref{app:conflict}), the deletion behind the margin target removes a violating event for 80.9/83.8\% of requests, about 17 points above the 64.0/67.2\% expected from history composition (Figure~\ref{fig:conflict-analysis}(c)). The excess shrinks to 2.9/4.1 points when the deletion is selected under the generic request, and reverses to $-12.3/{-11.1}$ points when the target itself violates the constraint, so the selection depends on both the request and the target.
\vspace{-3mm}
\paragraph{AIMS still uses user history.}
Training without history (\emph{Request-only CE}) beats Continued CE
overall on this dataset and reaches a violation rate similar to AIMS on
conflict requests, but it loses Recall@10 on requests without conflicting
history, and AIMS exceeds it overall and in both groups
(Appendix~\ref{app:history-use}). \emph{History dropout}, which empties each training history with
probability 0.5, trails AIMS by 0.176/0.378 Recall@10 points
($p<0.05$ on both backbones) and gains far less on conflict requests (+0.143/+0.192 vs.\ +0.558/+0.874;
Table~\ref{tab:history-use}). Swapping in other users' matched histories
lowers Recall@10 similarly for AIMS and Continued CE but only about half as
much for History dropout, so AIMS still uses each user's own history.

%% file: sections/2_related.tex
\vspace{-2mm}
\section{Related Work}
\label{sec:related_work}
\vspace{-2mm}
Personalized search asks when to rely on the user's
history~\citep{dou2007personalized,ai2019zam}, and \citet{lu2024controllable}
train LLM recommenders with category-based instructions and reinforcement
learning to follow stated intentions that conflict with past preferences.
CFT and GDRT improve history use via counterfactual logit comparisons
and group-robust training~\citep{zhang2024causality,wang2026context};
CauseRec, STEAM, and ReLLa synthesize, correct, or select the input
history~\citep{zhang2021causerec,lin2023steam,lin2024rella}; and
retrieval-augmented generation trains retrievers and inference-time
filters on label-dependent utility~\citep{izacard2023atlas,wang2023filco}.
LETTER, LTRGR, S-DPO, and OneRec use ranking or preference
losses~\citep{wang2024letter,li2024ltrgr,chen2024sdpo,deng2025onerec},
and Margin-MSE distills teacher score
margins~\citep{hofstatter2020marginmse}. AIMS instead turns single-event
deletions into request-specific margin targets while keeping complete
histories in training and inference, penalizing shortfalls via the
competitor only (Appendix~\ref{app:related_work}).

%% file: sections/6_conclusion.tex
\vspace{-2mm}
\section{Conclusion}
\label{sec:conclusion}
\looseness=-1
\vspace{-2mm}
We documented influence--utility misalignment, showed that target-score gains need not improve rankings, and proposed AIMS, which turns request-specific history deletions into counterfactual margin targets learned on complete histories through an asymmetric auxiliary loss, leaving inference unchanged. Across six backbones and three datasets, AIMS achieves the best Recall@10 and NDCG@10 in every setting, improving Recall@10 over matched Continued CE by 4.0--10.9\% (relative), with larger gains under instruction--history conflict that dropping history during training does not reproduce. The deletions behind its margin targets preferentially remove history that violates the request.

%% file: sections/appendix.tex
\providecommand{\TBD}[1]{\textbf{[#1]}}

\section{Protection-Set Statistics}
\label{app:protection}

Table~\ref{tab:protection} summarizes the requests that receive the
auxiliary supervision, for the two backbones used in the ablations. The
protection set is defined on training requests, which the SFT checkpoint
fits better than held-out requests. Most protected requests have at least
one eligible deletion, and almost all of these activate the hinge at
initialization, so the initial activation status is insensitive to
$\varepsilon$ near its default. A median gain of about 0.25 nats multiplies
the target-to-competitor likelihood ratio by about 1.3. Because the ACS
hinge (Section~\ref{sec:acs}) is linear in the margin shortfall, a larger
gain prolongs the activation of a request but does not scale its gradient. Appendix~\ref{app:placebo} compares these gains with gains from a random-item placebo.

\begin{table}[h]
    \centering
    \caption{\textbf{Protection-set statistics.} Share of training requests
    in the protection set $\mathcal P$, share of protected requests with an
    eligible deletion ($g_i>0$) and with an active hinge at initialization
    ($g_i>\varepsilon$), and quantiles of the positive gains (nats).}
    \label{tab:protection}
    \fontsize{8}{9.4}\selectfont
    \setlength{\tabcolsep}{4pt}
    \begin{tabular*}{0.9\linewidth}{@{\extracolsep{\fill}}lrrrrrrr@{}}
        \toprule
        & $|\mathcal P|$ & $g_i>0$ & $g_i>\varepsilon$
        & \multicolumn{4}{c}{Gain $g_i$ (nats)} \\
        \cmidrule(lr){5-8}
        Backbone & (\% of train) & (\% of $\mathcal P$) & (\% of $\mathcal P$)
                 & P25 & Median & P75 & P90 \\
        \midrule
        Llama-3.1-8B & 18.1 & 88.2 & 87.8 & 0.125 & 0.234 & 0.388 & 0.687 \\
        Gemma-3-12B  & 30.6 & 85.7 & 85.1 & 0.150 & 0.256 & 0.415 & 0.693 \\
        \bottomrule
    \end{tabular*}
\end{table}

\section{Experimental Details}
\label{app:setup}

\subsection{Training Pipeline}

\paragraph{Semantic identifiers.}
In all experiments, the decoder predicts three SID tokens, each from a
level-specific vocabulary of 2{,}048 codes. For the industrial dataset, we
use the first three levels of the platform's production semantic
identifiers, so that all datasets share the same SID format. For Qilin and
KuaiSearch-Lite, we train a separate three-level RQ-VAE over
Qwen3-Embedding-0.6B representations of the item text. SID mappings are
dataset-specific and fixed across all methods and backbones within each
dataset, and the 6{,}144 SID tokens are added to each backbone's
vocabulary.

\paragraph{Training stages.}
For each dataset and backbone, the base model is first trained to generate
item SIDs and then fine-tuned with CE on (request, history, target)
instances, which gives $\theta_0$. Continued CE, all baselines, and AIMS
start from $\theta_0$ with the same instances, batch size, and number of
steps.

\paragraph{Compute.}
All experiments were run on a multi-node cluster with eight NVIDIA A100 80GB GPUs per node.
\subsection{Datasets}

\paragraph{Industrial dataset.}
The industrial dataset is sampled from anonymized production logs of a
large-scale content platform. Each request is a natural-language user
query, the target is the item the user clicked for that request, and the
history is the user's 30 most recent interactions before the request,
represented by their SIDs. Training uses three consecutive days of logs and
evaluation a single later day, separated by a one-day gap.

\paragraph{Qilin.}
Qilin~\citep{chen2025qilin} contains app-level sessions with real Chinese
queries, the notes exposed for each request, click feedback, and up to 20
recently clicked notes. We use its search subset. For each request, the
query is $q$, the released pre-request click sequence is $\mathcal H$, and
$\mathcal Y$ is the set of distinct exposed notes that received a click. We
retain requests with $\mathcal Y\neq\emptyset$, expand each
$\mathbf y\in\mathcal Y$ into one training instance that keeps the request
identifier, and compute Recall@$k$ and NDCG@$k$ at the request level over
all positives in $\mathcal Y$. We keep the official test split (6{,}192
requests) and divide the 44{,}024 released training requests into 39{,}931
training and 4{,}093 development requests with a deterministic
user-disjoint hash split.

\paragraph{KuaiSearch-Lite.}
KuaiSearch \citep{li2026kuaisearch} contains real e-commerce search logs from the Kuaishou platform, with authentic queries, the items exposed for each request, click and purchase feedback, and item metadata including titles, brands, sellers, and three-level category hierarchies. We use KuaiSearch-Lite, the released subset that keeps the complete search and interaction records of 102,086 randomly sampled users over one month (555,553 queries). For each request, the query is $q$, $\mathcal{H}$ contains up to the 20 most recent clicked items provided by the ranking table, and $\mathcal{Y}$ is the set of distinct clicked items. Requests are filtered, expanded, and evaluated as on Qilin. Following the benchmark's temporal protocol, we use the last day of logs as the test split and partition the remaining requests into user-disjoint training and development sets, which yields 257,780 / 28,781 / 8,456 click-positive training / development / test requests.

\subsection{Evaluation Protocols}
 
\paragraph{Full-catalog retrieval.}
On the industrial dataset, every method runs SID-constrained beam search
(beam size 100) over the full catalog, which returns a ranked list of 100
unique valid SIDs per request; Recall@$k$ and NDCG@$k$ use the top $k$ of
this list. Recall@$k$ is the fraction of test requests whose target appears
in the top $k$, and NDCG@$k$ uses binary relevance with one target per
request.
 
\paragraph{Candidate-pool ranking.}
On the public benchmarks, each candidate in the pool of a request is scored
by the full-sequence SID log-probability of Eq.~(\ref{eq:score}) under
teacher forcing, on the complete request and history, and the pool is
sorted; the released order breaks exact ties. All clicked items of a
request are relevant: Recall@$k$ is the fraction of them in the top $k$,
and NDCG@$k$ uses binary gains with the ideal ranking placing all of them
first. Sampled-candidate metrics can deviate from full-ranking
metrics~\citep{krichene2020sampled}, so we use them only to compare methods
on identical pools; absolute values are not comparable across datasets.

 \paragraph{Pool extension.}
  Some released impression pools contain fewer candidates than the evaluation
  cutoff, making Recall@10 uninformative. We therefore deterministically extend
  pools containing fewer than 40 items with distinct items sampled from
  training-split impressions, while retaining all logged candidates and leaving
  larger pools unchanged. Sampling is seeded by the request identifier and does
  not use click labels; existing candidates, including all clicked items, are
  excluded from sampling. This rule is fixed across splits, and every method is
  evaluated on identical pools.

\paragraph{AIMS with candidate pools.}
Each clicked item of a request forms one training instance. For an
instance, the initial list of Section~3.1 is the pool sorted by the scores
of $\theta_0$ with the other clicked items of the same request removed: the
instance enters the protection set when its target is among the ten
highest-scored candidates of this list, and its competitor is the eleventh
candidate. Inference is unchanged for all methods.

\paragraph{Significance testing.}
All significance statements use a paired user-cluster bootstrap with 10,000 resamples: users are resampled with replacement, all test
requests of a user are kept together, every method is evaluated on the same
resamples, and a difference is called significant ($p<0.05$) when its
two-sided 95\% percentile interval excludes zero.
 
 \paragraph{Baselines.}
All baselines start from $\theta_0$ and share AIMS's SIDs,
positive training instances, training schedule, and evaluation
protocol. Hyperparameters are tuned on the same validation split.
Continued CE performs additional CE-only training.
CFT~\citep{zhang2024causality} adds a counterfactual CE loss
on the difference between full-history and history-absent
vocabulary logits at teacher-forced target positions.
LETTER (RG)~\citep{wang2024letter} applies its temperature-scaled
ranking-guided generation loss to our shared SIDs.
LTRGR~\citep{li2024ltrgr} combines generation and margin-based
ranking losses using SID sequence scores over model-retrieved
candidates.
S-DPO~\citep{chen2024sdpo} applies a multi-negative softmax
preference loss to reference-relative SID sequence
log-probabilities, with $\theta_0$ as the frozen reference.
  
\subsection{Hyperparameters}
\label{app:hparams}

Table~\ref{tab:hparams} lists all hyperparameters. All backbones on the
same dataset use the same values. Test splits are never used for
hyperparameter selection. Single-event deletion, one competitor,
$R=10$, and beam size 100 are fixed in advance.

\begin{table}[h]
      \centering
      \caption{\textbf{Hyperparameters.}}
      \label{tab:hparams}
      \fontsize{8}{9.4}\selectfont
      \setlength{\tabcolsep}{3.5pt}
      \begin{tabular*}{\linewidth}{@{\extracolsep{\fill}}lccc@{}}
          \toprule
          & Industrial & Qilin & KuaiSearch-Lite \\
          \midrule
          Tuning / LoRA
              & Full  & Full  & Full  \\
          Optimizer / learning rate
              & AdamW / $10^{-5}$
              & AdamW / $2{\times}10^{-6}$
              & AdamW / $2{\times}10^{-6}$ \\
          Global batch / updates
              & 128 / 2{,}500
              & 8 / 39{,}806
              & 32 / 37{,}268 \\
          Maximum history length
              & 30 & 20 & 20 \\
          Acceptance threshold $\tau$ (nats)
              & $10^{-4}$ & $10^{-4}$ & $10^{-4}$ \\
          Tolerance $\varepsilon$ (nats)
              & $0.01$ & $0.01$ & $0.01$ \\
          Loss weight $\lambda$
              & $1.0$ & $1.0$ & $1.0$ \\
          \bottomrule
      \end{tabular*}
  \end{table}

\subsection{Ablation variants}
\label{app:variants}

 \paragraph{Difficulty-based gain controls.}
\label{app:difficulty-controls}

Both controls replace the original request--gain assignment with a
difficulty-based assignment and use frozen targets
$T_i=m_i^0+\widehat g_i$, while retaining AIMS's protection set,
competitors, and training objective.

For \emph{D-free}, we define difficulty as $d_i=-m_i^0$ and compute
its empirical midrank percentile within the training protection set:
\[
    r_i =
    \frac{
        |\{j\in\mathcal P:d_j<d_i\}|
        +\tfrac12|\{j\in\mathcal P:d_j=d_i\}|
    }{|\mathcal P|}.
\]
We set $\widehat g_i=a r_i$, so requests with smaller initial
margins receive larger gains. We select
$a\in\{0,0.25,0.5,1,2,4\}$ nats by validation Recall@10 on
Llama-3.1-8B and share the selected value across backbones.
The percentiles and gains are fixed before continued training;
their construction uses no deleted-history scores or AIMS gains.

For \emph{D-cal}, we sort all AIMS gains, including zeros, in
ascending order and assign them to protected requests ordered
from easiest to hardest by $-m_i^0$, breaking difficulty ties
by request index. This preserves the marginal gain distribution
and the distribution of initial hinge losses
$[\widehat g_i-\varepsilon]_+$, but may change which requests
start active.

\paragraph{Generic-request gains.}
\label{app:gen}
Generic-matched recomputes the deletion gains under the generic request
$q^{\mathrm G}$ of Section~\ref{sec:ium_utility}, which omits the
specific intent. Eq.~(\ref{eq:clean_gains}) is evaluated with
$q^{\mathrm G}$ in place of $q_i$ in every score, so both gains compare
generic deleted and generic complete histories, and
Eqs.~(\ref{eq:aims_eligible}) and~(\ref{eq:clean_teacher}) then give a
generic gain $g_i^{\mathrm G}$. The protection set, competitors, and
initial margins $m_i^0$ remain those of the factual request. We rank the
protected requests by $g_i^{\mathrm G}$, breaking ties with a random key
fixed before training, and assign them the sorted AIMS gains, including
zeros, so that a larger generic gain receives a larger AIMS gain. The
frozen targets $T_i=m_i^0+\widehat g_i$ thus keep the AIMS gain
distribution, initial hinge losses, and number of initially active
requests, but change which requests receive each gain. The student uses
the factual request, complete history, and training configuration of
AIMS. These targets form a synthetic control and need not equal any
generic deletion-view margin.

\paragraph{Deletion-selection rules.}
\label{app:selection-rules}
\emph{Uplift-first} and \emph{Influence-first} keep the deletion views, the eligibility test of Eq.~(\ref{eq:aims_eligible}), the protection set, competitors, and training objective of AIMS, and change only which deletion defines $g_i$.
Uplift-first chooses the deletion with the largest target uplift $\Delta_{y,i}(S)$, the deletion audited in Figure~\ref{fig:instruction_ium}(c).
Influence-first chooses the deletion with the largest influence, the Jensen--Shannon divergence of Section~\ref{sec:ium_influence} recomputed on the deletion views; this choice uses no labels.
If the chosen deletion $S$ is eligible, $g_i=\Delta_{m,i}(S)$; otherwise $g_i=0$.
Because AIMS takes the largest margin gain over eligible deletions, both rules give $g_i\le g_i^{\mathrm{AIMS}}$ for every request.

\section{Additional Results}
\label{app:results}

\subsection{All cutoffs on the public benchmarks}
\label{app:cutoffs}
Tables~\ref{tab:qilin-full} and~\ref{tab:kuaisearch-full} report all
cutoffs for the public benchmarks in the layout of Table~\ref{tab:internal-main-results}.
 
\begin{table}[h]
    \centering
    \caption{\textbf{Performance (\%) on Qilin} under candidate-pool
    ranking. R@$k$ and N@$k$ denote Recall@$k$ and NDCG@$k$.}
    \label{tab:qilin-full}
    \fontsize{8}{9.4}\selectfont
    \setlength{\tabcolsep}{3pt}
    \renewcommand{\arraystretch}{0.94}
    \begin{tabular*}{\linewidth}{@{\extracolsep{\fill}}llrrrrrrr@{}}
        \toprule
        & & \multicolumn{4}{c}{Recall (\%) $\uparrow$} & \multicolumn{3}{c}{NDCG (\%) $\uparrow$} \\
        \cmidrule(lr){3-6}\cmidrule(lr){7-9}
        Model & Method & R@1 & R@5 & R@10 & R@20 & N@5 & N@10 & N@20 \\
        \midrule
        \multirow{6}{*}{Llama-3.1-8B}
        & Continued CE  & 4.72 & 21.07 & 35.48 & 60.72 & 16.80 & 22.92 & 31.11 \\
        & CFT           & 4.85 & 21.15 & 35.52 & 60.55 & 16.97 & 23.04 & 31.16 \\
        & LETTER (RG)   & 5.04 & 21.89 & 36.70 & 61.83 & 17.61 & 23.97 & 32.11 \\
        & LTRGR         & 5.31 & 21.48 & 36.13 & 61.12 & 17.22 & 23.45 & 31.63 \\
        & S-DPO         & 4.96 & 21.42 & 35.87 & 61.26 & 17.46 & 23.49 & 31.80 \\
        & AIMS (ours)   & 5.36 & 22.36 & 37.61 & 62.31 & 18.06 & 24.41 & 32.62 \\
        \midrule
        \multirow{6}{*}{Llama-3.1-70B}
        & Continued CE  & 6.27 & 27.17 & 43.47 & 64.35 & 21.58 & 28.47 & 35.28 \\
        & CFT           & 6.38 & 27.12 & 43.35 & 64.02 & 22.03 & 28.83 & 35.61 \\
        & LETTER (RG)   & 6.99 & 28.76 & 45.08 & 65.68 & 23.42 & 30.34 & 36.97 \\
        & LTRGR         & 6.60 & 27.98 & 44.30 & 65.11 & 22.29 & 29.11 & 35.98 \\
        & S-DPO         & 6.57 & 28.11 & 44.19 & 64.99 & 22.68 & 29.50 & 36.35 \\
        & AIMS (ours)   & 7.05 & 28.87 & 45.62 & 66.54 & 23.99 & 30.90 & 37.70 \\
        \midrule
        \multirow{6}{*}{Gemma-3-4B}
        & Continued CE  & 3.72 & 17.14 & 30.74 & 57.16 & 13.79 & 19.38 & 28.11 \\
        & CFT           & 3.81 & 16.93 & 30.63 & 57.04 & 13.76 & 19.55 & 28.37 \\
        & LETTER (RG)   & 3.89 & 17.51 & 31.45 & 58.11 & 14.07 & 19.91 & 28.65 \\
        & LTRGR         & 4.23 & 17.39 & 31.16 & 57.70 & 14.15 & 19.97 & 28.86 \\
        & S-DPO         & 3.94 & 17.88 & 31.94 & 58.28 & 14.73 & 20.71 & 29.05 \\
        & AIMS (ours)   & 4.18 & 18.52 & 32.79 & 59.20 & 14.80 & 20.95 & 29.53 \\
        \midrule
        \multirow{6}{*}{Gemma-3-12B}
        & Continued CE  & 5.18 & 21.04 & 36.29 & 61.11 & 17.30 & 23.68 & 31.74 \\
        & CFT           & 5.06 & 21.31 & 36.80 & 61.56 & 17.29 & 23.67 & 31.68 \\
        & LETTER (RG)   & 5.73 & 22.23 & 37.74 & 62.57 & 18.06 & 24.62 & 32.78 \\
        & LTRGR         & 5.35 & 21.64 & 37.09 & 61.89 & 17.69 & 24.15 & 32.18 \\
        & S-DPO         & 5.62 & 22.01 & 37.64 & 62.34 & 17.96 & 24.54 & 32.77 \\
        & AIMS (ours)   & 5.90 & 22.79 & 38.57 & 63.18 & 18.45 & 25.10 & 33.02 \\
        \midrule
        \multirow{6}{*}{Qwen3-8B}
        & Continued CE  & 5.16 & 22.36 & 37.45 & 62.68 & 18.20 & 24.74 & 32.88 \\
        & CFT           & 5.24 & 22.31 & 37.62 & 62.79 & 18.02 & 24.39 & 32.46 \\
        & LETTER (RG)   & 5.48 & 23.71 & 38.63 & 63.62 & 18.97 & 25.32 & 33.46 \\
        & LTRGR         & 5.53 & 23.23 & 38.41 & 63.54 & 18.79 & 25.12 & 33.55 \\
        & S-DPO         & 5.21 & 22.74 & 37.80 & 63.07 & 18.35 & 24.74 & 33.10 \\
        & AIMS (ours)   & 5.61 & 23.84 & 39.10 & 64.24 & 19.41 & 25.88 & 34.10 \\
        \midrule
        \multirow{6}{*}{Qwen3-14B}
        & Continued CE  & 5.91 & 24.12 & 39.81 & 64.11 & 19.65 & 26.24 & 34.00 \\
        & CFT           & 6.12 & 24.58 & 40.91 & 64.67 & 20.22 & 27.07 & 34.83 \\
        & LETTER (RG)   & 6.35 & 25.14 & 41.35 & 65.29 & 21.00 & 27.75 & 35.37 \\
        & LTRGR         & 6.34 & 25.06 & 41.42 & 65.48 & 20.59 & 27.43 & 35.48 \\
        & S-DPO         & 6.62 & 24.67 & 40.97 & 64.96 & 20.37 & 27.23 & 35.18 \\
        & AIMS (ours)   & 6.58 & 25.72 & 42.06 & 66.17 & 21.06 & 27.92 & 35.86 \\
        \bottomrule
    \end{tabular*}
\end{table}

\begin{table}[h]
    \centering
    \caption{\textbf{Performance (\%) on KuaiSearch-Lite} under candidate-pool
    ranking. R@$k$ and N@$k$ denote Recall@$k$ and NDCG@$k$.}
    \label{tab:kuaisearch-full}
    \fontsize{8}{9.4}\selectfont
    \setlength{\tabcolsep}{3pt}
    \renewcommand{\arraystretch}{0.94}
    \begin{tabular*}{\linewidth}{@{\extracolsep{\fill}}llrrrrrrr@{}}
        \toprule
        & & \multicolumn{4}{c}{Recall (\%) $\uparrow$} & \multicolumn{3}{c}{NDCG (\%) $\uparrow$} \\
        \cmidrule(lr){3-6}\cmidrule(lr){7-9}
        Model & Method & R@1 & R@5 & R@10 & R@20 & N@5 & N@10 & N@20 \\
        \midrule
        \multirow{6}{*}{Llama-3.1-8B}        & Continued CE  & 3.74 & 15.31 & 26.36 & 46.57 & 12.73 & 17.38 & 24.08 \\        & CFT           & 3.80 & 15.36 & 26.41 & 46.72 & 12.52 & 17.18 & 23.91 \\        & LETTER (RG)   & 3.87 & 15.81 & 26.87 & 47.11 & 12.91 & 17.58 & 24.15 \\        & LTRGR         & 4.07 & 15.86 & 27.21 & 47.46 & 12.98 & 17.75 & 24.38 \\        & S-DPO         & 3.77 & 15.48 & 26.54 & 46.88 & 12.50 & 17.00 & 23.76 \\        & AIMS (ours)   & 4.10 & 15.90 & 27.70 & 48.08 & 13.00 & 18.12 & 24.72 \\        \midrule        \multirow{6}{*}{Llama-3.1-70B}        & Continued CE  & 5.02 & 19.41 & 31.80 & 53.28 & 16.12 & 21.31 & 28.35 \\        & CFT           & 5.09 & 19.60 & 32.11 & 53.56 & 16.33 & 21.50 & 28.49 \\        & LETTER (RG)   & 5.33 & 20.67 & 33.43 & 54.94 & 16.92 & 22.26 & 29.38 \\        & LTRGR         & 5.42 & 20.32 & 33.16 & 54.72 & 16.67 & 22.14 & 29.19 \\        & S-DPO         & 5.22 & 20.09 & 32.71 & 54.27 & 16.44 & 21.60 & 28.79 \\        & AIMS (ours)   & 5.49 & 20.61 & 34.02 & 55.49 & 16.94 & 22.55 & 29.64 \\        \midrule        \multirow{6}{*}{Gemma-3-4B}        & Continued CE  & 3.19 & 12.95 & 23.11 & 42.28 & 10.61 & 14.91 & 21.30 \\        & CFT           & 3.22 & 13.18 & 23.35 & 42.53 & 10.84 & 15.14 & 21.30 \\        & LETTER (RG)   & 3.42 & 13.47 & 23.86 & 43.11 & 11.23 & 15.59 & 21.81 \\        & LTRGR         & 3.45 & 13.51 & 23.79 & 42.94 & 11.13 & 15.47 & 21.63 \\        & S-DPO         & 3.31 & 13.25 & 23.40 & 42.62 & 10.66 & 14.92 & 21.10 \\        & AIMS (ours)   & 3.58 & 14.04 & 24.75 & 44.36 & 11.65 & 16.15 & 22.48 \\        \midrule        \multirow{6}{*}{Gemma-3-12B}        & Continued CE  & 4.24 & 16.31 & 28.04 & 48.75 & 13.82 & 18.74 & 25.50 \\        & CFT           & 4.20 & 16.17 & 27.91 & 48.48 & 13.15 & 18.01 & 24.83 \\        & LETTER (RG)   & 4.43 & 16.55 & 28.80 & 49.09 & 13.66 & 18.71 & 25.38 \\        & LTRGR         & 4.55 & 16.84 & 28.39 & 49.54 & 13.90 & 18.98 & 25.83 \\        & S-DPO         & 4.48 & 16.69 & 28.53 & 49.27 & 13.72 & 18.72 & 25.34 \\        & AIMS (ours)   & 4.73 & 17.51 & 29.97 & 51.10 & 14.57 & 19.82 & 26.72 \\        \midrule        \multirow{6}{*}{Qwen3-8B}        & Continued CE  & 3.46 & 13.69 & 23.94 & 43.76 & 11.40 & 15.70 & 22.14 \\        & CFT           & 3.61 & 13.97 & 24.42 & 44.22 & 11.57 & 15.90 & 22.35 \\        & LETTER (RG)   & 3.86 & 14.62 & 25.40 & 45.47 & 12.21 & 16.71 & 23.31 \\        & LTRGR         & 3.91 & 14.68 & 25.33 & 45.31 & 12.21 & 16.69 & 23.37 \\        & S-DPO         & 3.74 & 14.51 & 25.18 & 45.12 & 12.07 & 16.51 & 22.93 \\        & AIMS (ours)   & 4.08 & 15.37 & 26.56 & 46.88 & 12.79 & 17.53 & 24.17 \\        \midrule        \multirow{6}{*}{Qwen3-14B}        & Continued CE  & 4.00 & 15.84 & 27.15 & 47.88 & 13.22 & 17.88 & 24.81 \\        & CFT           & 3.97 & 15.78 & 27.03 & 47.63 & 13.13 & 17.84 & 24.50 \\        & LETTER (RG)   & 4.28 & 16.48 & 28.13 & 49.02 & 13.62 & 18.56 & 25.44 \\        & LTRGR         & 4.22 & 16.31 & 27.82 & 48.61 & 13.39 & 18.19 & 25.02 \\        & S-DPO         & 4.31 & 16.44 & 28.05 & 48.92 & 13.56 & 18.49 & 25.40 \\        & AIMS (ours)   & 4.43 & 16.96 & 28.78 & 49.83 & 14.03 & 19.03 & 26.02 \\        \bottomrule
    \end{tabular*}
\end{table}
 
\subsection{Training seeds}

Continued CE, the strongest baseline of each backbone, and AIMS are each
trained with the same three seeds from the same $\theta_0$. The seed affects only the randomness of continued training. The cached
competitors and margins of AIMS are computed once from $\theta_0$ and are
identical across seeds. Table~\ref{tab:seeds} reports means, standard deviations, and paired
differences over seeds. All other results in the paper use one of these
seeds, the same for every experiment.

\begin{table}[t]
\centering
\caption{\textbf{Run-to-run variation.} Recall@10 (\%) over three training seeds,
reported as mean$_{\pm\text{SD}}$. The baseline is the strongest baseline of each
backbone by Recall@10 in the main results. $\Delta$~columns give the mean and SD
of the per-seed differences of AIMS, which is higher in all three seeds in every
comparison.}
\label{tab:seeds}
\footnotesize
\setlength{\tabcolsep}{3.5pt}
\begin{tabular*}{\linewidth}{@{\extracolsep{\fill}}llccccc@{}}
\toprule
& & \multicolumn{3}{c}{Recall@10 (\%)} & \multicolumn{2}{c}{$\Delta$ of AIMS} \\
\cmidrule(lr){3-5}\cmidrule(l){6-7}
Backbone & Baseline & Cont.\ CE & Baseline & AIMS & vs.\ CE & vs.\ baseline \\
\midrule
\multicolumn{7}{@{}l}{\emph{Industrial}} \\
Llama-3.1-8B & LETTER (RG) & $7.003_{\pm0.091}$  & $7.214_{\pm0.077}$  & $7.309_{\pm0.078}$  & $+0.306_{\pm0.048}$ & $+0.095_{\pm0.058}$ \\
Gemma-3-12B & LETTER (RG) & $7.955_{\pm0.032}$ & $8.499_{\pm0.040}$ & $8.582_{\pm0.052}$ & $+0.627_{\pm0.036}$ & $+0.083_{\pm0.027}$ \\
\addlinespace[2pt]
\multicolumn{7}{@{}l}{\emph{Qilin}} \\
Llama-3.1-8B & LETTER (RG) & $35.496_{\pm0.031}$ & $36.591_{\pm0.097}$ & $37.641_{\pm0.039}$ & $+2.145_{\pm0.013}$ & $+1.050_{\pm0.132}$ \\
Gemma-3-12B  & LETTER (RG) & $36.301_{\pm0.057}$ & $37.709_{\pm0.032}$ & $38.622_{\pm0.096}$ & $+2.321_{\pm0.045}$ & $+0.913_{\pm0.098}$ \\
\addlinespace[2pt]
\multicolumn{7}{@{}l}{\emph{KuaiSearch-Lite}} \\
Llama-3.1-8B & LTRGR       & $26.358_{\pm0.017}$ & $27.210_{\pm0.041}$ & $27.673_{\pm0.029}$ & $+1.315_{\pm0.022}$ & $+0.463_{\pm0.034}$ \\
Gemma-3-12B  & LETTER (RG) & $28.041_{\pm0.013}$ & $28.803_{\pm0.032}$ & $30.005_{\pm0.059}$ & $+1.964_{\pm0.048}$ & $+1.202_{\pm0.033}$ \\
\bottomrule
\end{tabular*}
\end{table}

\subsection{Deletion budget and competitor set}
\label{app:sensitivity}
 
Table~\ref{tab:sensitivity} varies the deletion budget and the competitor
set around the defaults ($K=1$, $M=1$). For $K>1$, a
non-stopping greedy search evaluates, in each round, all one-event
extensions of the previous round's raw-margin-gain maximizer, even when the
best intermediate gain is non-positive. The same eligibility conditions are
then applied, and the eligible set with the largest margin gain is
selected. At $K=3$, \emph{Conservative} selection instead takes the
smallest eligible set that attains at least a fraction $\eta = 0.9$ of the maximum gain. With
$M>1$, \emph{Pair-mean} averages ACS losses over the first $M$ excluded
candidates, and \emph{Set-max} compares against the highest-scoring
candidate in this fixed set. Larger budgets give small, backbone-dependent
changes with up to $2.9\times$ as many scoring views. Pair-mean changes are
small and inconsistent, and Set-max degrades as $M$ grows on both backbones.
 
\begin{table}[!htbp]
    \centering
    \caption{\textbf{Sensitivity to deletion and competitor choices.}
    Recall@10 and NDCG@10 (\%).
    Views/req. reports reference-model scoring views averaged across
    backbones, and Rel. gives the ratio to the default.
    Neither quantity represents training time.
    Dashes denote unreported counts.}
    \label{tab:sensitivity}
    \label{tab:deletion-budget}
    \label{tab:competitor-set}
 
    \fontsize{8}{9.2}\selectfont
    \setlength{\tabcolsep}{4pt}
    \renewcommand{\arraystretch}{0.92}
    \setlength{\aboverulesep}{0.20ex}
    \setlength{\belowrulesep}{0.20ex}
    \setlength{\cmidrulesep}{0.10ex}
 
    \begin{tabular*}{\linewidth}{
        @{\extracolsep{\fill}}llrrrrrr@{}
    }
        \toprule
        & &
        \multicolumn{2}{c}{Llama-3.1-8B}
        & \multicolumn{2}{c}{Gemma-3-12B}
        & \multicolumn{2}{c}{Scoring cost} \\
        \cmidrule(lr){3-4}
        \cmidrule(lr){5-6}
        \cmidrule(lr){7-8}
 
        Design & Setting
        & R@10 & N@10
        & R@10 & N@10
        & Views/req. & Rel. \\
        \midrule
 
        Default & $K=1$, $M=1$
        & 7.223 & 4.629
        & 8.609 & 5.598
        & 29.97 & $1.00\times$ \\
 
        \midrule
 
        \multicolumn{8}{@{}l}{
            \emph{A. Deletion budget and search policy}
        } \\
 
        Greedy & $K=2$
        & 7.230 & 4.581
        & 8.611 & 5.619
        & 58.94 & $1.97\times$ \\
 
        Greedy & $K=3$
        & 7.270 & 4.701
        & 8.578 & 5.659
        & 86.91 & $2.90\times$ \\
 
        Conservative & $K=3$, $\eta = 0.9$
        & 7.291 & 4.707
        & 8.609 & 5.609
        & 86.91 & $2.90\times$ \\
 
        \addlinespace[1.5pt]
 
        \multicolumn{8}{@{}l}{
            \emph{B. Competitor-set size and aggregation}
        } \\
 
        Pair-mean & $M=3$
        & 7.216 & 4.644
        & 8.641 & 5.595
        & --- & --- \\
 
        Pair-mean & $M=5$
        & 7.239 & 4.649
        & 8.594 & 5.598
        & --- & --- \\
 
        Set-max & $M=3$
        & 7.181 & 4.585
        & 8.581 & 5.584
        & --- & --- \\
 
        Set-max & $M=5$
        & 7.090 & 4.541
        & 8.491 & 5.568
        & --- & --- \\
 
        \bottomrule
    \end{tabular*}
\end{table}

\subsection{History use}
\label{app:history-use}

Request-only CE continues CE training from $\theta_0$ with the history
field left empty, using the same training requests, targets, and schedule
as Continued CE. History dropout uses the same setup but empties the
history of each training instance with probability 0.5, drawn independently
per instance from a stream fixed by the training seed, so it lies halfway
between Continued CE ($p=0$) and Request-only CE ($p=1$); we fix $p=0.5$ in
advance rather than tuning it, and inference uses the complete history.
For the history swap, we replace each request's history
with three histories of other users that match its request category,
history length, and coarse category composition. The replacements are
chosen without targets or model outputs and shared by all methods, and the
three results are averaged per request. Groups follow
Appendix~\ref{app:conflict}. Table~\ref{tab:history-use} reports the
results.

\begin{table}[h]
\centering
\caption{\textbf{History use on the industrial dataset.} Overall R@10 and
N@10 (\%) are computed on the full test set. $\Delta$R@10 is the change from
Continued CE on requests without conflicting history ($\kappa_i = 0$) and on
conflict requests ($\kappa_i \geq 0.5$). FAIL is the observed-violation rate (\%)
on conflict requests. Swap drop is the decrease in R@10 or N@10 when each
history is replaced with matched histories of other users, on requests
without conflicting history. Hist.\ dropout empties each training history
with probability 0.5. Dashes mark entries that do not apply:
$\Delta$R@10 for Continued CE, against which the changes are measured, and the swap drop for
Request-only CE, which uses no history.}
\label{tab:history-use}
\fontsize{8}{9.4}\selectfont
\setlength{\tabcolsep}{3pt}
\begin{tabular*}{\linewidth}{@{\extracolsep{\fill}}llrrrrrrr@{}}
\toprule
& & \multicolumn{2}{c}{Overall} & \multicolumn{1}{c}{No conflict}
  & \multicolumn{2}{c}{Conflict} & \multicolumn{2}{c}{Swap drop} \\
\cmidrule(lr){3-4}\cmidrule(lr){5-5}\cmidrule(lr){6-7}\cmidrule(lr){8-9}
Backbone & Method & R@10 & N@10 & $\Delta$R@10 & $\Delta$R@10 & FAIL
  & R@10 & N@10 \\
\midrule
\multirow{4}{*}{Llama-3.1-8B}
 & Continued CE    & 6.902 & 4.460 & ---    & ---    & 29.0 & 0.247 & 0.143 \\
 & Request-only CE & 6.984 & 4.493 & $-0.417$ & $+0.186$ & 20.5 & --- & --- \\
 & Hist.\ dropout  & 7.047 & 4.523 & $-0.192$ & $+0.143$ & 22.8 & 0.132 & 0.076 \\
 & AIMS            & 7.223 & 4.629 & $+0.080$ & $+0.558$ & 21.6 & 0.276 & 0.158 \\
\midrule
\multirow{4}{*}{Gemma-3-12B}
 & Continued CE    & 7.945 & 5.141 & ---    & ---    & 25.4 & 0.325 & 0.208 \\
 & Request-only CE & 8.106 & 5.239 & $-0.553$ & $+0.241$ & 18.6 & --- & --- \\
 & Hist.\ dropout  & 8.231 & 5.326 & $-0.268$ & $+0.192$ & 19.7 & 0.161 & 0.099 \\
 & AIMS            & 8.609 & 5.598 & $+0.120$ & $+0.874$ & 18.3 & 0.294 & 0.183 \\
\bottomrule
\end{tabular*}
\end{table}

\subsection{Placebo interventions and selection rules}
\label{app:placebo}

\paragraph{Protocol.} For each protected request of the industrial dataset, we replace each single-event deletion with a random-item placebo that inserts one sampled item at the $j$-th history position, for $j = 1, \dots, n_i$, keeping all original events and their order. Each view therefore contains $n_i + 1$ events; because nearly all industrial histories have the maximum length of 30, most views contain 31 events, one more than the longest training history. Inserted items are sampled from the training-set frequency of history items, excluding the user's history, the target, and the competitor, with seeds fixed by request and position. We score the target and competitor under each view with the frozen reference model $\theta_0$ and apply Eqs.~(6)--(8) with the same eligibility rule and threshold $\tau$, which gives a placebo gain $g_i^{\mathrm{pl}}$ over the same number of views as the deletion audit.

\begin{table}[t]
\centering
\caption{Deletion, selection-rule, and placebo gains on the industrial dataset. Deletion is AIMS. Uplift-first and Influence-first choose one deletion per request, with the largest target uplift or the largest influence (Appendix~\ref{app:selection-rules}), and set $g_i=0$ if it is ineligible; Deletion and Placebo take the best eligible view of each request. Placebo views insert one random item at each history position instead of deleting an event, with the same protected requests, reference model, eligibility rule, and number of views; views of full histories contain 31 events. Shares are percentages of $\mathcal{P}$; quantiles are over positive gains (nats), as in Table~\ref{tab:protection}. $\rho_{\mathrm{all}}$ is the Spearman correlation with the deletion gains of AIMS over all protected requests, including zero gains.}
\label{tab:placebo}
\small
\begin{tabular}{llrrrrrrr}
\toprule
Backbone & Gain source
& \shortstack{$g>0$\\(\% of $\mathcal{P}$)}
& \shortstack{$g>\varepsilon$\\(\% of $\mathcal{P}$)}
& P25 & Median & P75 & P90
& $\rho_{\mathrm{all}}$ \\
\midrule
Llama-3.1-8B & Deletion & 88.2 & 87.8 & 0.125 & 0.234 & 0.388 & 0.687 & -- \\
             & Placebo  & 66.4 & 45.1 & 0.007 & 0.018 & 0.039 & 0.074 & 0.064 \\
                          & Uplift-first    & 75.9 & 71.4 & 0.049 & 0.123 & 0.258 & 0.478 & 0.670 \\
             & Influence-first & 28.8 & 25.9 & 0.030 & 0.080 & 0.193 & 0.371 & 0.197 \\
\addlinespace
Gemma-3-12B  & Deletion & 85.7 & 85.1 & 0.150 & 0.256 & 0.415 & 0.693 & -- \\
             & Placebo  & 69.7 & 49.8 & 0.008 & 0.023 & 0.048 & 0.091 & 0.091 \\
                          & Uplift-first    & 66.1 & 61.7 & 0.046 & 0.120 & 0.249 & 0.435 & 0.559 \\
             & Influence-first & 25.9 & 23.4 & 0.033 & 0.090 & 0.205 & 0.372 & 0.195 \\
\bottomrule
\end{tabular}
\end{table}
\paragraph{Results.} Table~\ref{tab:placebo} reports the results. Placebo perturbations yield an eligible view for 66.4\% (Llama) and 69.7\% (Gemma) of protected requests, versus 88.2\% and 85.7\% for deletions, and gains above $\varepsilon$ for 45.1\% and 49.8\%, versus 87.8\% and 85.1\%. Their positive gains are an order of magnitude smaller (median 0.018 and 0.023 nats vs.\ 0.234 and 0.256). The lower quartile of positive deletion gains (0.125 and 0.150) exceeds the 90th percentile of positive placebo gains (0.074 and 0.091). Across protected requests, the Spearman correlation between deletion and placebo gains is 0.064 and 0.091. Insertions and deletions are not symmetric perturbations: they differ in history length, event positions, and content. The comparison therefore shows only that the tested random insertion does not produce gains comparable to deletion; it does not isolate the cause of the difference. Appendix~\ref{app:conflict} examines directly whether the deleted events conflict with the request.

\paragraph{Selection rules.}
The two selection rules of Appendix~\ref{app:selection-rules} leave fewer eligible deletions and smaller gains than AIMS (Table~\ref{tab:placebo}).
Unlike AIMS and the placebo, which take the best eligible view of each request, each rule commits to one deletion.
The largest-uplift deletion is eligible for 75.9/66.1\% of protected requests (Llama/Gemma), versus 88.2/85.7\% for AIMS; the requests it loses are those whose largest-uplift deletion raises the target score but not the margin, the failure audited in Figure~\ref{fig:instruction_ium}(c).
The most influential deletion is eligible for only 28.8/25.9\%, and its gains are weakly correlated with the AIMS gains ($\rho_{\mathrm{all}}=0.197/0.195$).
Positive gains are also smaller under both rules (medians 0.123/0.120 and 0.080/0.090 versus 0.234/0.256 nats), although $g_i\le g_i^{\mathrm{AIMS}}$ per request does not by itself imply lower quantiles over positive gains.

\subsection{Retention--recovery decomposition}
\label{app:retention}
On the industrial dataset, each test request has one target, and all methods are evaluated on the same split, defined by the top-10 decisions of $\theta_0$. The Recall@10 difference between two methods therefore decomposes exactly as $\Delta\mathrm{R@10} = p_0\,\Delta\mathrm{Ret} + (1-p_0)\,\Delta\mathrm{Rec}$, where $\Delta$ denotes AIMS minus Continued CE and $p_0$ is the test hit rate of $\theta_0$ (6.70/7.70\% for Llama-3.1-8B/Gemma-3-12B). With the retention and recovery rates of Figure~4, the share of the gain due to higher recovery, $(1-p_0)\,\Delta\mathrm{Rec}/\Delta\mathrm{R@10}$, is 68/85\%.

\subsection{Signed label-free attribution}\label{app:proxies}
For each audit request of Section~2.2, let $\hat{\mathbf{y}}$ be the top-1 SID decoded by $\theta_0$ on the complete history. The utility $u_{j,\hat{\mathbf{y}}}$, a standard removal-based attribution for the model's own prediction~\citep{covert2021removing}, is computed on the same masked views as $\rho_j$ and $U_j$ and evaluated with the same two statistics. Because $u_{j,\hat{\mathbf{y}}}=U_j$ when $\hat{\mathbf{y}}=\mathbf{y}$, Table~\ref{tab:proxies} also reports requests with $\hat{\mathbf{y}}\neq\mathbf{y}$. On these requests, $u_{j,\hat{\mathbf{y}}}$ is negatively correlated with $U_j$ on both backbones, and its positive values over all Gemma requests come from requests with $\hat{\mathbf{y}}=\mathbf{y}$. Whether $\hat{\mathbf{y}}$ is the target is unknown at inference, so neither attribution reliably identifies target-supporting events.

\begin{table}[t]
\centering
\caption{Label-free attributions versus signed target utility on the 1,600 audit requests under factual requests. Spearman is the mean within-history correlation with $U_j$. $\Delta$Supp.\ is the change in supportive-event rate (percentage points) when selecting the event with the largest attribution, relative to uniform selection. Brackets give 95\% paired user-cluster bootstrap intervals. When $\hat{\mathbf{y}}=\mathbf{y}$, $u_{j,\hat{\mathbf{y}}}=U_j$.}
\label{tab:proxies}
\small
\begin{tabular}{lrlcc}
\toprule
Requests & $N$ & Attribution & Spearman & $\Delta$Supp. \\
\midrule
\multicolumn{5}{l}{\textit{Llama-3.1-8B}} \\
All & 1,600 & JS influence $\rho_j$ & $0.045\;[0.035,\,0.055]$ & $+0.24\;[-2.14,\,2.62]$ \\
All & 1,600 & Signed $u_{j,\hat{\mathbf{y}}}$ & $0.019\;[0.003,\,0.036]$ & $-1.89\;[-4.35,\,0.54]$ \\
$\hat{\mathbf{y}}\neq\mathbf{y}$ & 1,488 & JS influence $\rho_j$ & $0.046\;[0.036,\,0.057]$ & $-0.27\;[-2.68,\,2.14]$ \\
$\hat{\mathbf{y}}\neq\mathbf{y}$ & 1,488 & Signed $u_{j,\hat{\mathbf{y}}}$ & $-0.055\;[-0.066,\,-0.044]$ & $-5.78\;[-8.25,\,-3.32]$ \\
\midrule
\multicolumn{5}{l}{\textit{Gemma-3-12B}} \\
All & 1,600 & JS influence $\rho_j$ & $0.002\;[-0.008,\,0.012]$ & $+0.42\;[-1.95,\,2.74]$ \\
All & 1,600 & Signed $u_{j,\hat{\mathbf{y}}}$ & $0.066\;[0.047,\,0.085]$ & $+3.61\;[1.23,\,6.01]$ \\
$\hat{\mathbf{y}}\neq\mathbf{y}$ & 1,424 & JS influence $\rho_j$ & $0.003\;[-0.008,\,0.014]$ & $+0.98\;[-1.52,\,3.44]$ \\
$\hat{\mathbf{y}}\neq\mathbf{y}$ & 1,424 & Signed $u_{j,\hat{\mathbf{y}}}$ & $-0.050\;[-0.061,\,-0.038]$ & $-2.46\;[-4.96,\,0.04]$ \\
\bottomrule
\end{tabular}
\end{table}

\section{Instruction--History Conflict Evaluation Protocol}
\label{app:conflict}
We analyze the industrial test requests (25\% of the test set) that state an
explicit positive content-category constraint, extracted by a deterministic
parser that uses neither targets nor model outputs. Violations and history
conflict are only defined for such requests.

\paragraph{Constraint and history labels.}
Item categories are obtained from the platform's frozen item-category
metadata and bound to three-level semantic item identifiers (SIDs). Item
evaluation likewise uses neither targets nor model outputs, and the same
labels are shared by all methods and backbones.

Only positive history events strictly preceding the request are eligible.
For the conjunction of the request's explicit constraints, an eligible
history item is \textsc{pass} if every constraint is satisfied,
\textsc{fail} if at least one constraint is demonstrably violated, and
\textsc{unknown} if no violation is established but the available
attributes are insufficient to verify all constraints.  Let
$P_i,F_i,U_i$ denote these counts for request $i$.  A request enters the
severity analysis only if
\begin{equation}
  P_i+F_i \geq 3
  \quad\text{and}\quad
  \frac{P_i+F_i}{P_i+F_i+U_i}\geq 0.8.
  \label{eq:conflict-coverage}
\end{equation}
We define the continuous history-conflict severity
\begin{equation}
  \kappa_i=\frac{F_i}{P_i+F_i}
  \label{eq:conflict-severity}
\end{equation}
and stratify eligible requests into $\kappa_i=0$, $0<\kappa_i<0.5$,
$0.5\leq\kappa_i<0.8$, and $0.8\leq\kappa_i\leq1$.  The 3.8\% of requests
that do not meet the evaluable-history coverage condition are excluded.

\paragraph{Output labels and metrics.}
Exactly ten ranks are materialized for every request.  Each slot receives
the same \textsc{pass}, \textsc{fail}, or \textsc{unknown} label as a
history item.  It is \textsc{invalid} if the rank is missing, repeats an
earlier SID, or cannot be mapped into the frozen candidate and
item-attribute snapshot.  For a stratum containing $N_b$ requests, the
observed violation rate is
\begin{equation}
  \mathrm{FAIL}_b=\frac{n_{b,\textsc{fail}}}{10N_b}.
  \label{eq:severity-violation}
\end{equation}
\textsc{unknown} and \textsc{invalid} slots remain in the fixed denominator
but do not enter the numerator and are not treated as satisfying the
constraint.  We define
$\Delta\mathrm{FAIL}_b=\mathrm{FAIL}_b^{\mathrm{method}}-\mathrm{FAIL}_b^{\mathrm{CE}}$,
so negative values indicate fewer violations, and report the AIMS relative
reduction $-\Delta\mathrm{FAIL}_b/\mathrm{FAIL}_b^{\mathrm{CE}}$.

\begin{table}[t]
  \centering
 \caption{\textbf{Performance by history-conflict severity.} Request shares
are percentages of the constraint-bearing test requests, and FAIL rates are
percentages of output slots labeled as confirmed violations. Changes are
percentage points relative to Continued CE, with 95\% paired user-cluster
bootstrap intervals in brackets; $\Delta$R@10 intervals are reported for
AIMS only. Negative $\Delta$FAIL means fewer confirmed violations, and
Rel.\ is $-\Delta$FAIL divided by the CE FAIL rate.}
  \label{tab:conflict-dose-response}
  \fontsize{8}{9.4}\selectfont
  \setlength{\tabcolsep}{3pt}
  \renewcommand{\arraystretch}{0.96}
  \begin{tabular*}{\linewidth}{@{\extracolsep{\fill}}llrrlrrr@{}}
    \toprule
    Backbone & $\kappa_i$ stratum & Share & CE FAIL & Method
      & $\Delta$FAIL [95\% CI] & Rel.\ (\%) & $\Delta$R@10 [95\% CI] \\
    \midrule
    Llama-3.1-8B
      & $0$         &  3.9 &  8.4 & D-cal & $-0.4\ [-1.75,\ 1.01]$ &  4.8 & $+0.027$ \\
      &             &      &      & AIMS  & $-0.6\ [-2.09,\ 0.91]$ &  7.1 & $+0.080\ [-0.18,\ 0.32]$ \\
      & $(0,0.5)$   &  8.6 & 14.2 & D-cal & $-1.0\ [-2.01,\ 0.05]$ &  7.0 & $+0.091$ \\
      &             &      &      & AIMS  & $-2.1\ [-3.23,\ -1.03]$ & 14.8 & $+0.212\ [0.01,\ 0.38]$ \\
      & $[0.5,0.8)$ & 31.1 & 22.5 & D-cal & $-2.1\ [-2.69,\ -1.54]$ &  9.3 & $+0.187$ \\
      &             &      &      & AIMS  & $-5.3\ [-5.96,\ -4.71]$ & 23.6 & $+0.397\ [0.29,\ 0.50]$ \\
      & $[0.8,1]$   & 52.6 & 32.8 & D-cal & $-3.5\ [-3.92,\ -3.10]$ & 10.7 & $+0.308$ \\
      &             &      &      & AIMS  & $-8.6\ [-9.07,\ -8.19]$ & 26.2 & $+0.653\ [0.58,\ 0.73]$ \\
    \midrule
    Gemma-3-12B
      & $0$         &  3.9 &  7.6 & D-cal & $-0.3\ [-1.57,\ 0.93]$ &  3.9 & $+0.043$ \\
      &             &      &      & AIMS  & $-0.5\ [-1.92,\ 0.88]$ &  6.6 & $+0.120\ [-0.15,\ 0.37]$ \\
      & $(0,0.5)$   &  8.6 & 12.7 & D-cal & $-0.9\ [-1.84,\ -0.04]$ &  7.1 & $+0.159$ \\
      &             &      &      & AIMS  & $-1.8\ [-2.86,\ -0.86]$ & 14.2 & $+0.343\ [0.13,\ 0.51]$ \\
      & $[0.5,0.8)$ & 31.1 & 19.8 & D-cal & $-2.0\ [-2.56,\ -1.51]$ & 10.1 & $+0.283$ \\
      &             &      &      & AIMS  & $-4.9\ [-5.53,\ -4.35]$ & 24.7 & $+0.623\ [0.50,\ 0.72]$ \\
      & $[0.8,1]$   & 52.6 & 28.7 & D-cal & $-3.4\ [-3.77,\ -3.05]$ & 11.8 & $+0.499$ \\
      &             &      &      & AIMS  & $-8.4\ [-8.82,\ -8.04]$ & 29.3 & $+1.022\ [0.94,\ 1.10]$ \\
    \bottomrule
  \end{tabular*}
\end{table}

\paragraph{Uncertainty.}
We use 10,000 paired user-cluster bootstrap resamples with seed 154.
Users are sampled with replacement, all requests from a sampled user are
kept together, and every method is evaluated on the identical resample.
The reported confidence limits are the empirical 2.5th and 97.5th
percentiles; they are not centered or symmetrized around the
original-sample estimate.

\paragraph{Severity trend.}
Table~\ref{tab:conflict-dose-response} reports the results by stratum, and Figure~\ref{fig:conflict-analysis}(a,b) plots them. Continued CE's FAIL rate increases monotonically with $\kappa_i$, indicating
that the frozen history statistic captures how strongly historical
evidence points away from the current constraint.  AIMS's absolute and
relative violation reductions also increase monotonically.  From
$\kappa_i=0$ to $\kappa_i\geq0.8$, the absolute reduction grows from 0.6 to
8.6 points on Llama and from 0.5 to 8.4 points on Gemma; the highest- and
lowest-severity intervals do not overlap.  The point estimates of the
Recall@10 gains are non-decreasing over the same strata, so there is no
evidence in these estimates that the compliance improvement trades off
against target retrieval.  In point estimates, D-cal improves over
Continued CE in every stratum but remains below AIMS, with the gap
widening as conflict severity increases. These are observational strata. The monotone association does not by
itself identify conflict severity as the causal source of the gains.

\paragraph{Labels of selected deletions.}
We test whether the deletions selected by AIMS remove history events that violate the request's constraint.
We apply the same parser and frozen metadata to the protected training requests of the industrial dataset and label the history event removed by each selected deletion (Eq.~(\ref{eq:clean_teacher})); these labels are used only for this analysis.
We keep requests with an explicit category constraint that meet Eq.~(\ref{eq:conflict-coverage}), have a \textsc{pass} target, and contain both \textsc{pass} and \textsc{fail} history events ($0<\kappa_i<1$).
If the selection ignored the labels, it would remove a \textsc{fail} event with probability $F_i/n_i$, where $n_i$ counts all single-event deletions of request~$i$ (\textsc{unknown} events stay in the denominator).
Table~\ref{tab:sel-labels} and Figure~\ref{fig:conflict-analysis}(c) compare the observed share with the mean of this probability.
To compare with the generic request of Section~\ref{sec:ium_utility}, the first two rows use the requests with an eligible deletion under both requests, so they share $N$ and the expected share.
The selected deletion removes a \textsc{fail} event 16.8/16.6 points more often than expected.
Under the generic request, with labels still taken from the factual constraint, the excess falls to 2.9/4.1 points.
When the target itself violates the constraint, \textsc{fail} events are removed 12.3/11.1 points less often than expected.
The labels do not enter AIMS; they only describe which events its selection removes.

\begin{table}[t]
  \centering
  \caption{\textbf{Constraint labels of the deletions selected by AIMS} on the industrial training protection set. Rows use protected requests with an explicit category constraint that meet Eq.~(\ref{eq:conflict-coverage}) and contain both \textsc{pass} and \textsc{fail} history events; the target is \textsc{pass} except in the Target \textsc{fail} rows. All and Generic request use the requests with an eligible deletion under both the factual and the generic request, so they share $N$ and Exp. Obs.: share of requests whose selected deletion removes a \textsc{fail} event. Exp.: mean of $F_i/n_i$, the share expected if the selection ignored the labels. Enrich.: Obs.\ minus Exp.\ in percentage points, computed before rounding, with 95\% user-cluster bootstrap intervals in brackets. Generic request: the deletion is selected under the generic request of Section~\ref{sec:ium_utility}, as for Generic-matched (Appendix~\ref{app:gen}); labels still refer to the factual constraint.}
  \label{tab:sel-labels}
  \fontsize{8}{9.4}\selectfont
  \setlength{\tabcolsep}{3pt}
  \renewcommand{\arraystretch}{0.96}
  \begin{tabular*}{\linewidth}{@{\extracolsep{\fill}}llrrrr@{}}
    \toprule
    Backbone & Requests & $N$ & Obs.\ (\%) & Exp.\ (\%) & Enrich.\ (pp) [95\% CI] \\
    \midrule
    Llama-3.1-8B
      & All                  & 2,674 & 80.85 & 64.04 & $+16.82\ [15.43,\ 18.24]$ \\
      & Generic request      & 2,674 & 66.94 & 64.04 & $+2.90\ [1.22,\ 4.55]$ \\
      & Target \textsc{fail} &   275 & 51.64 & 63.95 & $-12.32\ [-17.92,\ -6.86]$ \\
    \midrule
    Gemma-3-12B
      & All                  & 4,259 & 83.75 & 67.17 & $+16.58\ [15.47,\ 17.67]$ \\
      & Generic request      & 4,259 & 71.28 & 67.17 & $+4.11\ [2.82,\ 5.41]$ \\
      & Target \textsc{fail} &   702 & 55.70 & 66.78 & $-11.08\ [-14.49,\ -7.70]$ \\
    \bottomrule
  \end{tabular*}
\end{table}


\section{Extended Related Work}
\label{app:related_work}
 
This appendix expands Section~\ref{sec:related_work} and relates the
removal-based quantities of Section~\ref{sec:ium} to attribution methods.
 
\paragraph{Generative and instruction-guided recommendation.}
Generative and LLM-based recommendation has been studied through
text-to-text task formulations~\citep{geng2022p5, yin2024poisoningfederatedrecommendersystems}, efficient instruction
tuning~\citep{bao2023tallrec}, and semantic-identifier
generation~\citep{rajput2023generative,zheng2024lcrec,wang2024letter}.
In e-commerce search, OneSearch generates semantic identifiers from the
query together with short- and long-term user behavior sequences and
refines personalized ranking with a preference-aware reward
system~\citep{onesearch2025}.
Instruction-guided recommenders condition on explicit user
requests~\citep{zhang2025instructrec,lu2024controllable,xu2025iagent},
and LLM recommenders can underuse sequential
information~\citep{kim2025lost}. These works establish instruction
following and history use as objectives, whereas AIMS supplies
intervention-derived supervision for the target--competitor comparison
under a given request. Its asymmetry concerns auxiliary gradient routing,
not the asymmetric input--output prediction tasks of
LC-Rec~\citep{zheng2024lcrec}.
 
\paragraph{Request--history conflict.}
Studies of personalized web search found that personalization helps
queries with ambiguous intent but has little effect, or even hurts,
when the intent is clear~\citep{dou2007personalized,teevan2008personalize}.
Personalized product search combines the query with the user's
history~\citep{ai2017hem}; the zero attention model attends over the
user's history conditioned on the query and can attend to none of
it~\citep{ai2019zam}, and denoising attention has been proposed for
query-aware user modeling~\citep{bassani2024denoising}. Among LLM
recommenders, \citet{lu2024controllable} train the model to follow the
stated intention when it conflicts with preferences inferred from the
history, using control instructions synthesized from item categories
with labels from a sequential recommender, followed by reinforcement
learning; our requests are instead real user queries with observed
click targets. We therefore treat that pipeline as complementary to AIMS
rather than as a directly comparable baseline. In LLMs more broadly, models may follow either the
supplied context or their parametric knowledge when the two
conflict~\citep{longpre2021entity,xie2024adaptive}, context-aware
decoding contrasts predictions with and without the context at
inference~\citep{shi2024trusting}, and the instruction hierarchy trains
models to favor higher-priority instructions~\citep{wallace2024instruction}.
These works analyze or control how much a model relies on the history
or context, through its architecture, decoding, or training data.
AIMS instead evaluates individual history events offline under the
current request, using the observed target, and turns their effects
into training supervision, leaving the architecture and inference
unchanged.
 
\paragraph{Label-dependent context utility as supervision.}
Retrieval-augmented generation also derives supervision from
label-dependent estimates of context utility. Atlas trains its
retriever to predict how much each retrieved document improves the
likelihood of the output, and also evaluates a leave-one-out objective
based on how much the prediction degrades when one retrieved document
is removed~\citep{izacard2023atlas}. FILCO labels useful context spans
with label-dependent criteria, including string inclusion, lexical
overlap, and the ratio of gold-output probabilities with and without a
span as the supplied context; it then trains a filter that runs without
gold outputs at inference and supplies the filtered context to the
generator~\citep{wang2023filco}. Because filtering can also discard
relevant passages, \citet{yoran2024making} instead train the reader on
mixed relevant and irrelevant contexts. AIMS also measures signed
utility with the label: in our audit (Section~\ref{sec:ium}), the tested
label-free influence measure is not a reliable proxy for signed target
support. AIMS differs in what it intervenes on and how it uses the
result. It deletes individual events of a user's history under the
current request, and it turns their effects into a margin target for
the recommender itself rather than training a retriever or filter or
editing the input, so the recommender keeps the complete history during
training and inference. Training on the same deletion views instead is
less effective (Section~\ref{sec:ablations}). Every model in that comparison
uses the complete history at inference, so it concerns how deletions enter
training. A filter trained on our deletion labels and applied at inference
would instead add a selection step to deployment, and it would have to
predict target-dependent effects without the target (Table~\ref{tab:sel-labels}).
 
\paragraph{History synthesis, selection, and correction.}
CauseRec scores concepts relative to the next-item target and replaces
indispensable or dispensable concepts to construct contrastive
user-sequence representations~\citep{zhang2021causerec}. It therefore
provides a precedent for target-aware history transformations used
during training. STEAM learns keep, delete, and insert operations and
supplies corrected sequences to its recommender~\citep{lin2023steam}.
In candidate-conditioned click-through-rate prediction, ReLLa retrieves
semantically relevant historical interactions and mixes retrieved and
original views during tuning~\citep{lin2024rella}. In full-catalog generative
retrieval there is no candidate item at inference, so ReLLa would have to
retrieve history by the request, an inference-time selection step like the
filters discussed above. These approaches
improve recommendation through sequence synthesis, correction, or
selection. AIMS evaluates the contribution of historical evidence to
the current target--competitor comparison. An accepted deletion need
not identify an erroneous interaction: the recorded behavior may be
valid while contributing adversely under the present instruction.
 
\paragraph{Context and output debiasing in LLM recommenders.}
CFT combines standard likelihood training with a target-label objective
on the difference between full-history and history-absent
logits~\citep{zhang2024causality}. Its counterfactual signal is thus
label-directed, rather than an unsigned sensitivity objective.
GDRT groups target-token occurrences according to their predictability
without user history and applies group-distributionally robust
tuning~\citep{wang2026context}. This changes training emphasis across
groups, whereas AIMS constructs a margin target for an individual
request. SPRec addresses output bias through self-play: observed
interactions provide positive responses, previous model predictions
provide negatives, and SFT alternates with DPO~\citep{gao2025sprec}.
AIMS instead keeps its initially decoded competitor fixed throughout
continued training. At decoding time, D$^3$ addresses amplification and
homogeneity by removing length normalization associated with nearly
deterministic tokens and incorporating scores from an auxiliary
recommender~\citep{bao2024decoding}. AIMS retains the original
SID-constrained deployment decoder.
 
\paragraph{Ranking objectives and asymmetric negative supervision.}
Relative item preference is central to recommendation objectives such
as BPR~\citep{rendle2009bpr}. In generative retrieval, GR$^2$ incorporates
graded relevance through identifier design and constrained contrastive
learning~\citep{tang2024graded}. Among our baselines, LETTER (RG) applies
LETTER's temperature-scaled ranking-guided generation
loss~\citep{wang2024letter}, and LTRGR adds a learning-to-rank phase with
a rank loss over retrieved candidates~\citep{li2024ltrgr}. OneRec applies
iterative preference alignment to generated recommendation sessions,
using a reward model to select preferred and rejected
outputs~\citep{deng2025onerec}. For language-model alignment, DPO
optimizes response pairs using reference-relative sequence
probabilities, while SimPO uses length-normalized likelihoods and a
target reward margin~\citep{rafailov2023dpo,meng2024simpo}. S-DPO,
another baseline, extends DPO to multiple negatives for LM-based
recommenders~\citep{chen2024sdpo}. Unlikelihood training complements
likelihood maximization with penalties on undesirable
outputs~\citep{welleck2020unlikelihood}. These objectives shape the
ranked list, preferences between outputs, or the likelihood of
undesired outputs in general. AIMS instead derives its comparison
threshold from reference margins, strengthened by eligible history
interventions, and gates its auxiliary penalty by the student's margin
shortfall.
 
Recent ACPO likewise uses asymmetric preference optimization, retaining
the chosen-response gradient while adaptively scaling the
rejected-response gradient~\citep{huang2026acpo}. AIMS instead stops
the target branch of its auxiliary hinge and updates the competitor
branch when the constraint is active. Routing the auxiliary gradient
through both scores, as symmetric pairwise losses do, lowers Recall@10
relative to competitor-only routing (Section~\ref{sec:ablations}).
Analyses of finetuning dynamics and likelihood displacement show why
direct gradient choices must be distinguished from their effects on the
full predictive distribution~\citep{ren2025learning,razin2025displacement}.
In particular, shared parameters can still change target scores
indirectly despite the stopped auxiliary branch.
 
\paragraph{Transferring supervision and preserving recommendations.}
Learning with different teacher and student inputs connects AIMS to
generalized distillation~\citep{lopezpaz2016distillation} and context
distillation, which transfers benefits obtained from a teacher's
context into a model evaluated with a different
context~\citep{snell2022context}. More directly, Margin-MSE trains neural
rankers to match cached teacher score differences between document
pairs~\citep{hofstatter2020marginmse}. AIMS also transfers a scalar
comparison, but strengthens it with eligible history deletions and
treats it as a minimum desired margin with a tolerance, rather than
an equality target.
 
The focus on initially correct requests connects to the broader goal
of retaining existing capabilities during adaptation, as in Learning
without Forgetting~\citep{li2016forgetting}. AIMS operates within one
recommendation task, with the original training requests available.
Its reference supplies a selected target--competitor comparison,
which an eligible intervention can strengthen. This selective
constraint complements CE without requiring the student to reproduce
the reference's entire output distribution. A KL penalty toward the
reference model, as in the KL-regularized objective underlying
DPO~\citep{rafailov2023dpo}, is the distribution-level form of such
preservation, and our S-DPO baseline optimizes a preference loss derived
from this objective with $\theta_0$ as the frozen reference
(Appendix~\ref{app:setup}). R-comp, which sets $m_i^{\mathrm{cf}}=m_i^0$,
is the margin-level form of this preservation within AIMS: it improves Recall@10 over Continued CE by
+0.093/+0.336, and the request-specific gains account for the remaining
71\%/49\% of the improvement of AIMS (Section~\ref{sec:ablations}).
 
\paragraph{Attribution and counterfactual explanations.}
The audit of Section~\ref{sec:ium} measures how a prediction changes
when a history event is removed, which relates it to attribution.
Quantifying a feature's contribution through removal is a
well-established explanation strategy~\citep{covert2021removing}.
Additive attribution methods such as SHAP also assign signed
contributions to individual predictions~\citep{lundberg2017shap}.
Influence functions instead trace predictions to training examples
through approximate changes in model parameters~\citep{koh2017influence}.
ACCENT extends this approach to the score difference between a
recommended item and a replacement, searching for a small set of
historical training interactions whose removal would reverse the
recommendation~\citep{tran2021accent}. Its intervention concerns
estimated retraining effects, whereas AIMS directly evaluates input
deletions with a frozen model and uses their label-directed improvements
as supervision. Our diagnostic term \emph{influence} specifically denotes
target-free distributional sensitivity, distinct from the signed
influence-function quantities used in this literature. The utility
readout is a predictive contrast under a specified intervention,
rather than an identified causal effect of a user's behavior in
the world.